\documentclass{article}
\usepackage{graphicx}

\def\3nab{\tilde{\nabla}}

\def\hsp5{\hspace{5mm}}

\def\case#1/#2{\textstyle\frac{#1}{#2}}

\def\be {\begin{equation}}
\def\ee {\end{equation}}
\def\bea {\begin{eqnarray}}
\def\eea {\end{eqnarray}}

\def\case#1/#2{\textstyle\frac{#1}{#2} }

\def\cqg{{\em Class. Quantum Grav.\/} }
\def\grg{{\em Gen. Rel. Grav.\/} }
\def\prd{{\em Phys. Rev.\/} D }
\def\prl{{\em Phys. Rev. Lett.\/} }
\def\apj{{\em Astrophys. J.\/} }
\def\jmp{{\em J. Math. Phys.\/} }

\title{100 Years of General Relativity}
\author{George F R Ellis\\
	Mathematics Department, University of Cape Town}

\begin{document}
\maketitle
\setcounter{tocdepth}{2}
\tableofcontents

\begin{abstract}
	\textit{This is 
 Chapter 1 in 
{\rm General Relativity and Gravitation: A Centennial Perspective}, Edited by  
Abhay Ashtekar (Editor in Chief), Beverly Berger, 
James Isenberg, Malcolm MacCallum. 
Publisher: Cambridge University Press
(June, 2015). It gives a survey of themes that have been developed during the 100 years of development of general relativity theory.}
\end{abstract}
This chapter aims to provide a broad historical overview of the major
developments in General Relativity Theory (`GR') after the theory had been
developed in its final form. It will not relate the well-documented story of
the discovery of the theory by Albert Einstein, but rather will consider the
spectacular growth of the subject as it developed into a mainstream branch
of physics, high energy astrophysics, and cosmology. Literally hundreds of
exact solutions\index{exact solutions} of the full non-linear field equations are now known,
despite their complexity \cite{SteKraMac03}. The most important ones are the
Schwarzschild\index{Schwarzschild solution} and Kerr\index{Kerr solution} solutions, determining the geometry of the solar
system and of black holes\index{black hole} (Section \ref{Schw}), and the Friedmann-Lema\^{\i}%
tre-Robertson-Walker%
\index{Friedmann-Lema{\i}tre-Robertson-Walker spacetimes|see{FLRW models}}%
\index{FLRW models} solutions, which are basic to cosmology (Section \ref%
{FLRW}). Perturbations of these solutions make them the key to astrophysical
applications.

Rather than tracing a historical story, this chapter is structured in terms
of key themes in the study and application of GR:

1. The study of dynamic geometry (Section \ref{Geom}) through development of
various technical tools, in particular the introduction of global methods,
resulting in global existence and uniqueness theorems%
\index{existence theorems}\index{uniqueness theorems} and singularity
theorems\index{singularity theorems}.

2. The study of the vacuum Schwarzschild solution and its application to the
Solar system (Section \ref{Schw}), giving very accurate tests of general
relativity, and underlying the crucial role of GR in the accuracy of useful
GPSindex{GPS} systems.

3. The understanding of gravitational collapse%
\index{gravitational collapse} and the nature of Black Holes\index{black hole}
(Section \ref{BH}), with major applications in astrophysics, in particular
as regards quasi-stellar objects and active galactic nuclei.

4. The development of cosmological models (Section \ref{FLRW}), providing
the basis for our understandings of both the origin and evolution of the
universe as a whole, and of structure formation within it;

5. The study of gravitational lensing\index{gravitational lensing} and its astronomical applications,
including detection of dark matter\index{dark matter} (Section \ref{lens}).

6. Theoretical studies of gravitational waves\index{gravitational waves}, in particular resulting in
major developments in numerical relativity\index{numerical relativity} (Section \ref{gravwaves}), and
with development of gravitational wave observatories that have the potential
to become an essential tool in precision cosmology.\newline

This Chapter will not discuss quantum gravity, covered in 
Part Four.
It is of course impossible to refer to all relevant literature. I
have attempted to give the reader a judicious mix of path-breaking original
research articles, and good review articles. Ferreira has recently
discussed the historical development at greater length \cite{Fer14}.


\section{The study of dynamic geometry}

\label{Geom} This section deals with the study of dynamic geometry through
development of various technical tools, in particular the introduction of
global methods resulting in global existence and uniqueness theorems and
singularity theorems.%
\index{existence theorems}\index{uniqueness theorems}%
\index{singularity theorems}.

General relativity \cite{Ein15,MisThoWhe73,HawEll73,Wal84,Ste90}
heralded a new form of physical effect: geometry was no
longer seen as an eternal fixed entity, but as a dynamic physical variable.
Thus geometry became a key player in physics, rather than being a fixed
background for all that occurs. Accompanying this was the radical idea that
there was no gravitational force, rather that matter curves spacetime, and the
paths of freely moving particles are geodesics determined by the spacetime.
This concept of geometry as dynamically determined by its matter content
necessarily leads to the non-linearity of both the equations and the
physics. This results in the need for new methods of study of these
solutions; standard physics methods based on the assumption of linearity
will not work in general.

Tensor calculus\index{tensor calculus} is a key tool in general
relativity \cite{SynSch49,Sch54}. The \textbf{space-time geometry} is
represented on some specific 
averaging scale and determined by the\index{metric} \textbf{metric} $g_{ab}(x^{\mu })$.
The curvature tensor\index{curvature} $R_{abcd}$ is given by the Ricci
identities\index{Ricci identities} for
an arbitrary vector field $u^{a}$: 
\begin{equation}
u_{b;[cd]}=u^{a}R_{abcd}
\end{equation}%
where square brackets denote the skew part on the relevant indices, and a
semi-colon the covariant derivative. The curvature tensor plays a key role
in gravitation through its contractions, the Ricci tensor\index{Ricci tensor} $%
R_{ab}:=R_{\;acb}^{c}$ and Ricci scalar\index{Ricci scalar} $R:=R_{\;\;a}^{a}.$ The \textbf{%
matter} present determines the geometry, through \textbf{Einstein's
relativistic gravitational field equations}
(`EFE') \cite{Ein15}\index{Einstein equations} given by

\begin{equation}
G_{ab}\equiv R_{ab}-{{\textstyle{\frac{1}{2}}}}\,R\,g_{ab}={\kappa }%
T_{ab}-\Lambda \,g_{ab}\ .  \label{eq:efe}
\end{equation}%
Geometry in turn determines the motion of the matter because the \textbf{%
twice-contracted Bianchi identities}\index{Bianchi identities} guarantee the conservation of total
energy-momentum:\index{energy-momentum tensor!conservation} 
\begin{equation}
\nabla _{b}G^{ab}=0\hsp5\Rightarrow \hsp5\nabla _{b}T^{ab}=0\ ,
\label{eq:cons}
\end{equation}%
provided the \textbf{cosmological constant}\index{cosmological constant} $\Lambda $ satisfies the
relation $\nabla _{a}\Lambda =0$, i.e., it is constant in time and space. In
conjunction with suitable equations of state for the matter, represented by
the stress-energy tensor $T_{ab},$  equations (\ref{eq:efe}) determine the
combined dynamical evolution of the model and the matter in it, with (\ref%
{eq:cons}) acting as integrability conditions.

\subsection{Technical developments}

Because of the non-Euclidean geometry, coordinate freedom is a major feature
of the theory, leading to the desirability of using covariant equations that
are true in all coordinate systems if they are true in one. Because of the
non-linear nature of the field equations, it is desirable to use exact
methods and obtain exact solutions as far as possible, in order to not miss
phenomena that cannot be investigated through linearized versions of the
equations. A series of technical developments facilitated study of these
non-linear equations.

\subsubsection{Coordinate free methods and general bases\label{Coord_free}}

The first was the use of coordinate free methods, representing vector fields
as differential operators, and generic bases rather than just coordinates
bases. Thus one notes the differential geometry idea that tangent vectors
are best thought of as operators acting on functions \cite{HawEll73,EllMaaMac12}, thus $X=X^{i}\frac{\partial }{\partial x^{i}}\Rightarrow
X(f)=X^{i}\frac{\partial f}{\partial x^{i}},$ with a coordinate basis $%
e_{i}= $ $\frac{\partial }{\partial x^{i}}.$ Then a generic basis
$e_{a}$~($a=0,1,2,3$) is given by 
\[
e_{a}=\Lambda _{a}^{\;i}(x^{j})e_{i}=\Lambda _{a}^{\;i}(x^{j})\frac{\partial 
}{\partial x^{i}},\quad |\Lambda _{a}^{\;i}|\neq 0. 
\]%
There are three important aspects of any basis. First, the commutator
coefficients $\gamma _{\;bc}^{a}(x^{c})$ defined by%
\[
\gamma _{\;bc}^{a}e_{a}=[e_{b},e_{c}],\quad [X,Y]:=XY-YX. 
\]%
The basis $e_{a}$ is a coordinate basis iff\footnote{iff means ``if and only if''} $\gamma _{\;bc}^{a}=0.$ \
Second, the metric components\index{metric} $g_{ab}$ are defined by
\[
g_{ab}:=e_{a}.e_{b}=\Lambda _{a}^{\;i}\Lambda _{b}^{\;j}g_{ij} 
\]%
with inverses $g^{ab}$ determined by $g^{ab}g_{bc}=\delta _{c}^{a}.$ \
Indices are raised and lowered by $g_{ab}$ and $g^{bc}.$ The basis is a 
\textit{tetrad}\index{tetrads} basis if the $g_{ab}$ are constants. The two key forms of
tetrad are null tetrads with two real null vectors and two complex ones,
used for studying gravitational radiation\index{gravitational radiation}, where
\[
g_{ab}=\left( 
\begin{array}{cccc}
0 & -1 & 0 & 0 \\ 
-1 & 0 & 0 & 0 \\ 
0 & 0 & 0 & 1 \\ 
0 & 0 & 1 & 0%
\end{array}%
\right) , 
\]%
and orthonormal tetrads, used for studying fluid properties, where
\[
g_{ab}=\left( 
\begin{array}{cccc}
-1 & 0 & 0 & 0 \\ 
0 & 1 & 0 & 0 \\ 
0 & 0 & 1 & 0 \\ 
0 & 0 & 0 & 1%
\end{array}%
\right) . 
\]%
Third, there are the rotation coefficients $\Gamma _{\;bc}^{a}$
characterizing the covariant derivatives of the basis vectors, defined by 
\[
\nabla _{b}e_{c}=\Gamma _{\;bc}^{a}e_{a}. 
\]%
Using the standard assumptions of (1) metricity: writing $f_{,i}=\frac{%
\partial f}{\partial x^{i}}$ this is 
\[
\nabla _{e}g_{dc}=0\;\Longleftrightarrow \;g_{dc,e}\ =\Gamma _{dec}+\Gamma
_{ced} 
\]%
and (2) vanishing torsion: writing $f_{;ab}:=(f_{;a})_{;b}$, this is  
\[
f_{;ab}=f_{;ba}\;\forall f(x^{i})\;\Leftrightarrow \;\gamma
_{\;ab}^{c}=\Gamma _{\;ab}^{c}-\Gamma _{\;ba}^{c} 
\]%
one obtains (3) the generalized Christoffel
relations\index{Christoffel relations}%
\[
\Gamma _{ced}={{\textstyle{\frac{1}{2}}}}\,(g_{cd,e}+g_{ec,d}-g_{de,c})+{{%
\textstyle{\frac{1}{2}({\gamma }_{edc}+{\gamma }_{dec}-\gamma }}}_{ced}). 
\]%
The first term vanishes for a tetrad, and the second for a coordinate basis.
Tetrad bases\index{tetrads} were used in obtaining solutions by Levi-Civita in the 1920s.
Null tetrads\index{tetrads!null} were the basis of the Newman-Penrose formalism used primarily
to study gravitational radiation \cite{PenRin84}; they are closely related
to spinorial variables \cite{PenRin84}. Orthonormal
tetrads\index{tetrads!orthonormal}
 have been used
to study fluid models and Bianchi spacetimes\index{Bianchi models} \cite{Ell67,EllMac69}.
Dual 1-form relations, using exterior derivatives, have been used by many
workers (e.g. Bel, Debever, Misner, Kerr) to find exact solutions.

\subsubsection{Tensor symmetries and the volume element\label{sec:symmetries}%
}

Second was a realization of the importance of tensor symmetries, for example
separating a tensor $T_{ab}$ into its symmetric and skew symmetric parts,
and then separating the former into its trace and trace-free parts:%
\[
T_{ab}=T_{[ab]}+T_{(ab)},\;T_{(ab)}=T_{<ab>}+{{\textstyle{\frac{1}{4}}}}%
Tg_{ab},\;T_{<ab>}g^{ab}=0 
\]%
where $T_{[ab]}={{\textstyle{\frac{1}{2}}}}\,(T_{ab\ }-\;T_{ba}),$ $T_{(ab)}=%
{{\textstyle{\frac{1}{2}}}}\,(T_{ab\ }+\;T_{ba}),$ $T=T_{ab}g^{ab}.$ This
breaks the tensor $T_{ab}$ up into parts with different physical meanings. A
tensor equation implies equality of each part with the same symmetry, for
example 
\[
T_{ab}=W_{ab}\Leftrightarrow \left(
T_{[ab]}=W_{[ab]},\;T_{<ab>}=W_{<ab>},\;T\ =W\right) . 
\]%
This plays an important role in many studies. Similar decompositions occur
for more indices, for example $T_{[abc]}={{\textstyle{\frac{1}{6}}}}%
\,(T_{abc\ }+T_{bca\ }+T_{cab\ }-\;T_{acb}-T_{bac\ }-T_{cba\ })$. An
example is the curvature tensor symmetries%
\[
R_{abcd}=R_{[ab][cd]}=R_{cdab,}\;R_{a[bcd]}=0. 
\]

Arbitrarily large symmetric tensors can be split up in a similar way into
trace-free parts; this is significant for gravitational radiation studies 
\cite{Tho80} and kinetic theory, in particular studies of CMB anisotropies 
\cite{ChaLas99,LewChaLas00}. In the case of skew tensors, there is a
largest possible skew tensor, namely the volume element $\eta _{abcd}=\eta
_{\lbrack abcd]}$, which satisfies a key set of identities:%
\[
\eta _{abcd}\eta ^{efgh}=-4!\delta _{\lbrack a}^{[e}\delta _{b}^{f}\delta
_{c}^{g}\delta _{d]}^{h]} 
\]%
and others that follow by contraction. These symmetries characterize
invariant subspaces of a tensor product under the action of the linear group
and so embody an aspect of group representation theory: how to decompose
tensor representations into irreducible parts.

\subsubsection{Symmetry groups}
\index{symmetry groups}

Third was the systematization of the use of symmetry groups in studying
exact solutions. A symmetry is generated by a Killing vector field $\xi $, a
vector field that drags the metric into itself, and so gives a zero Lie
derivative for the metric tensor:
\[
L_{\xi }g_{ab}=0\;\Leftrightarrow \;\xi _{(a;b)}=0. 
\]%
Such a vector field satisfies the integrability condition 
\[
\xi _{a;bc}=R_{abcd}\xi ^{d} 
\]%
showing that a solution is determined by the values $\xi _{a|P}$ and $\xi
_{a;b|P}$ at a point P. The set of all Killing vectors form a Lie algebra
generating the symmetry group for the spacetime. The isotropy group of a
point is generated by the set of Killing vector fields vanishing at that
point. Exact solutions can be characterized by the group of symmetries,
together with a specification of the dimension and causal character of the
surfaces of transitivity of the group \cite{Ell67}. The most important
symmetries are when a spacetime is (a) static or stationary, (b)\
spherically symmetric, or (c) spatially homogeneous.

Killing vectors\index{Killing vector} give integration constants for geodesics\index{geodesic}: if $k^{a}$ is a
geodesic tangent vector, then $E:=\xi _{a}k^{a}$ is a constant along the
geodesic. They also give conserved vectors from the stress tensor: $%
J^{a}=T^{ab}\xi _{b}$ has vanishing divergence:  $J_{\;;a}^{a}=0.$

\subsubsection{Congruences of curves and geodesic deviation}
\index{geodesic deviation|(}

Fourth was the study of timelike and null congruences\index{congruences!timelike}\index{congruences!null} of curves (Synge,
Heckmann, Sch\"{u}cking, Ehlers, Kundt, Sachs,
Penrose) \cite{Syn37,Ehl61,Sac62}, leading to a realisation of the
importance of the 
geodesic deviation equation (GDE) \cite{Syn34} and its physical meaning in
terms of expressing tidal forces and gravitational
radiation\index{gravitational radiation}
\cite{Pir56,Pir57}). \newline
\ \qquad

The kinematical properties of null and timelike vector fields are
represented by their acceleration $a_{e},$ expansion ${\theta }$, shear $%
\sigma _{de}$, and rotation $\omega _{de}$ (which are equivalent to some of
the Ricci rotation coefficients for associated
tetrads \cite{NewPen62,Ell67}). They characterise the properties of
fluid flows (the timelike
case), hence are important in the dynamics of fluids, and of bundles of null
geodesics (the null case with $a_e=0$), and so are important in observations
in astronomy and cosmology.

The GDE determines the second rate of change of the deviation vectors for a
congruence of geodesics of arbitrary causal character, i.e., their relative
acceleration. Consider the normalised tangent vector field $V^{a}$ for such
a congruence, parametrised by an affine parameter $v$. Then $V^{a}:=\frac{%
dx^{a}(v)}{dv}\ ,\,\,\,V_{a}\,V^{a}:=\epsilon \ ,\,\,\,\,0={\frac{\delta
V^{a}}{\delta v}}=V^{b}\nabla _{b}V^{a}\ ,$ where $\epsilon =+\,1,\,0,\,-\,1$
if the geodesics are spacelike, null, or timelike, respectively, and we
define covariant derivation \emph{along\/} the geodesics by $\delta
T^{a..}{}_{b..}/\delta v:=V^{b}\nabla _{b}T^{a..}{}_{b..}$ for any tensor $%
T^{a..}{}_{b..}$. A deviation vector $\eta ^{a}:=dx^{a}(w)/dw$ for the
congruence, which can be thought of as linking pairs of neighbouring
geodesics in the congruence, commutes with $V^{a}$, so 
\begin{equation}
L_{V}\eta =0\Leftrightarrow {\frac{\delta \eta ^{a}}{\delta v}}=\eta
^{b}\nabla _{b}V^{a}\ .  \label{com}
\end{equation}%
It follows that their scalar product is constant along the geodesics: ${%
\frac{\delta (\eta _{a}V^{a})}{\delta v}}=0\,\,\,\,\Leftrightarrow
\,\,\,\,(\eta _{a}V^{a})=\mbox{const}\ .$ To simplify the relevant
equations, one can choose them orthogonal: $\eta _{a}\,V^{a}=0\ .$ The
general GDE takes the form 
\begin{equation}
\frac{\delta ^{2}\eta ^{a}}{\delta v^{2}}=-\,R^{a}\!_{bcd}\,V^{b}\,\eta
^{c}\,V^{d}\ ,  \label{gde}
\end{equation}%
This shows how spacetime curvature causes focussing or defocussing of
geodesics, and is the basic equation for gravitational lensing. The general
solution to this second-order differential equation along any geodesic $%
\gamma $ will have two arbitrary constants (corresponding to the different
congruences of geodesics that might have $\gamma $ as a member). There is a 
\emph{first integral\/} along any geodesic that relates the connecting
vectors for two \emph{different\/} congruences which have one central
geodesic curve (with affine parameter $v$) in common. This is $\eta
_{1}{}_{a}\,{\frac{\delta \eta _{2}}{\delta v}}^{a}-\eta _{2}{}_{a}\,{\frac{%
\delta \eta _{1}}{\delta v}}^{a}=\mbox{const}\ $ and is completely
independent of the curvature of the spacetime.

The trace of the geodesic deviation equation for timelike geodesics is the
Raychaudhuri equation \cite{Ray55}. In the case of timelike vectors\ $%
u^{a}=dx^{a}/d{\tau }$ it is the fundamental equation of gravitational
attraction for a fluid flow \cite{Ehl61,Ell71}:

\begin{equation}
\frac{{d\theta }}{{d\tau }}=-\frac{1}{3}\theta ^{2}-2(\omega ^{2}-\sigma
^{2})-{\textstyle{\frac{{\kappa }}{2}}}(\rho +3p)+\Lambda -a_{\,\,\,;b}^{b}.
\label{ray2}
\end{equation}%
In the case of null geodesics $k^{a}=dx^{a}/d\lambda $ diverging from a
source (so $\varepsilon =0,$ $a_{e}=0,~\omega =0$) it is the basic equation
of gravitational focusing of bundles of light rays \cite{HawEll73}:

\begin{equation}
\frac{d\theta }{d\lambda }+{\textstyle{\frac{1}{2}}}\theta ^{2}+\sigma
^{2}=-R_{ab}k^{a}k^{b}.  \label{obs1}
\end{equation}%
These equations play a key role in singularity
theorems\index{singularity theorems}.
\index{geodesic deviation|)}

\subsubsection{The conformal curvature tensor}
\index{conformal curvature|(}\index{Weyl tensor}

Fifth was a realisation, following on from this, that one could focus on the
full curvature tensor $R_{abcd}$ itself, and not just the Ricci tensor. The
curvature tensor $R_{abcd}$ is comprised of the Ricci tensor $R_{ab}$ and
the Weyl conformal curvature tensor $C_{abcd}$, given by

\begin{equation}
C_{abcd}:=R_{abcd}+{\textstyle{\frac{1}{2}}}%
(R_{ac}g_{bd}+R_{bd}g_{ac}-R_{ad}g_{bc}-R_{bc}g_{ad})-{\textstyle{\frac{1}{6}%
}}R(g_{ac}g_{bd}-g_{ad}g_{bc}).  \label{eq:weyl}
\end{equation}%
This has the same symmetries as the curvature tensor but in addition is
trace-free: $C_{\;acd}^{c}=0.$ The Ricci tensor\index{Ricci tensor} is determined pointwise by
the matter present through the field equations, but the Weyl tensor is not
so determined: rather it is fixed by matter elsewhere plus boundary
conditions. Its value at any point is determined by the Bianchi
identities\index{Bianchi identities}
\begin{equation}
R_{ab[cd;e]}=0  \label{Bianchi}
\end{equation}%
which are integrability conditions for the curvature tensor that must always
be satisfied. In 4 dimensions this gives both the divergence identities $%
\nabla ^{d}(R_{cd}-{\textstyle{\frac{1}{2}}}Rg_{cd})=0$ for the Ricci
tensor, which imply matter conservation (see (\ref{eq:cons})), and
divergence relations for $C_{abcd}:$ 
\begin{equation}
\;\nabla ^{d}C_{abcd}=\nabla _{\lbrack a}(-R_{b]c}+{\textstyle{\frac{1}{6}}}%
Rg_{b]c})\,.  \label{divR}
\end{equation}%
Substituting from the field equations (\ref{eq:efe}), matter tensor
derivatives are a source for the divergence of the Weyl tensor. Thus one can
think of the Weyl tensor as the free gravitational field (it is not
determined by the matter at a point), being generated by matter
inhomogeneities and then propagating to convey information on distant
gravitating matter to local systems. It will then affect local matter
behaviour through the geodesic deviation equation. Thus these are
Maxwell-like equations governing tidal forces and gravitational radiation
effects \cite{MaaBas98,EllMaaMac12}.

This geometrical and physical significance of the Weyl tensor has led to
studies of its algebraic structure (the Petrov
Classification)\index{Petrov classification} inter alia
by Petrov \cite{Pet54}, Pirani, Ehlers, Kundt, and by Penrose using a spinor
formalism \cite{Pen60,PenRin84}, and use of the Weyl tensor
components as auxiliary variables in studies of exact solutions,
gravitational radiation, and cosmology. One can search for vacuum solutions
of a particular Petrov type\index{exact solutions} \cite{SteKraMac03}, and relate asymptotic power
series at large distances to outgoing radiation conditions
\cite{Sac62,NewPen62,PenRin84}.
\index{conformal curvature|)}

\subsubsection{Junction Conditions\label{sec:join}}
\index{junction conditions}

Sixth, many solutions have different domains with different properties, for
example vacuum and fluid-filled. An important question then is how to join
two different such domains together without problems arising at the join. \
Lichnerowicz (using a coordinate choice) and Darmois (using coordinate-free
methods) showed how to join domains smoothly together, and Israel \cite%
{Isr66} showed how to assign properties to shock waves, boundary surfaces,
and thin shells that could lead to such junctions with a surface layer
occurring. The Darmois--Lichnerowicz case is included as the special
situation where there is no surface layer.

\subsubsection{Generation techniques}

Finally, a set of generation techniques were discovered that generated new
exact solutions from old ones, for example fluid solutions from vacuum
ones. These are discussed extensively in \cite{SteKraMac03}.


\subsection{Exact Theorems and Global Structure \label{global}}

The results mentioned so far are local in nature, but there has been a major
development of global results also (building on the local methods mentioned
above). This is covered in detail in Chapter 9 (`Global Behavior
 of Solutions to Einstein's Equations').


\subsubsection{Global properties and causality}

The first requirement is a careful use of coordinate charts as parts of
atlases that cover the whole spacetime considered and so avoid coordinate
singularities. This enables study of global topology\index{topology}, which may often not be
what was first expected, and is closely related to the causal structure of
the manifold. It is clear that closed timelike lines%
\index{closed timelike lines} can result if a
spacetime is closed in the timelike direction, but G\"{o}del showed that
closed timelike lines could occur for exact solutions of the field equations
for pressure-free matter that are simply connected \cite{God49}. This occurs
basically because, due to global rotation of the matter, light cones tip
over as one moves further from the origin. This paper led to an intensive
study of causation in curved spacetimes by Penrose, Carter, Geroch, Hawking,
and others. The field equations of classical general relativity do not
automatically prevent causality violation\index{causality violation}: so various causality
conditions have been
proposed as extra conditions to be imposed in addition to the Einstein
equations, the most physically relevant being stable causality (no closed
timelike lines exist even if the spacetime is perturbed) \cite{HawEll73}.
The global structure of examples such as the G\"{o}del universe and
Taub-NUT space were crucial in seeing the kinds of pathologies that can
occur.

\subsubsection{Conformal diagrams and horizons}

Studying the conformal structure of a spacetime is greatly facilitated by
using conformal diagrams\index{Penrose diagram}. Penrose pioneered this method \cite{Pen64} and
showed that one can rescale the conformal coordinates so that the boundary
of spacetime at infinite distance is represented at a finite coordinate
value, hence one can represent the entire spacetime and its boundary in this
way \cite{Pen64}. For example Minkowski space%
\index{Minkowski space!compactification} has null infinities $I_{-}$
and $I_{+}$ for incoming and outgoing null geodesics, an infinity $i_{0}$\
for spacelike geodesics, and past and future infinities $i_{-}$, $i_{+}$ for
timelike geodesics, and, perhaps surprisingly, the points $i_{0}$,
$i_{-}$ and
$i_{+}$ have to be identified. Penrose diagrams are now a standard tool in general
relativity studies, particularly in cosmology, where they make the structure
of particle horizons and visual horizons very clear, and in studying black
holes \cite{TipClaEll80}.

In the case of cosmology, there was much confusion about the nature of
horizons\index{horizons!cosmological}%
\index{horizons!event|see{event horizons}}
 until Rindler published a seminal paper that clarified the
confusion \cite{Rin56}. Penrose then showed that the nature of particle
horizons could be well understood by characterising the initial singularity
in cosmology as spacelike \cite{Pen64}. Nowadays they are widely used in
studying inflationary cosmology (but papers on inflation\index{inflation} often call the
Hubble radius the horizon, when it is not; the particle horizon is
non-locally defined). The key feature of black holes is event
horizons\index{event horizon},
which are null surfaces bounding the regions that can send information to
infinity from those which cannot; they occur when the future singularity is
spacelike. In the case of non-rotating black holes, their nature and
relation to the various spacetime domains (inside and outside the event
horizon), the existence of two spacelike singularities, and two separate
asymptotically flat exterior regions, is fully clarified by the associated
Penrose diagrams\index{Penrose diagram} (Section \ref{BH}). Carter developed the much more complex
such diagrams for the Reissner-Nordstrom charged solution and the Kerr
rotating black hole solution \cite{HawEll73}. Without these diagrams, it
would be very hard indeed to understand their global structure.

\subsubsection{Initial data, existence and uniqueness theorems}

Provided there are no closed timelike lines, existence of suitable initial
data on a spacelike surface S satisfying the initial value equations \cite%
{Ste38}, together with suitable equations of state for whatever matter may
be present, determines a unique solution for the Einstein equations within
the future domain of dependence of S\index{uniqueness theorems}: that is, the region of spacetime such
that all past timelike and null curves intersect S. Applied to the
gravitational equations, the spacetime developing from the initial data is
called the future Cauchy development\index{Cauchy problem!Cauchy development} of the data on
S \cite{HawEll73,Ren05}.
An important technical point is that one can prove
existence\index{existence theorems} of
timelike and null geodesics from S to every point in this domain of
dependence. The spacetime is called globally hyperbolic\index{global hyperbolicity} if the future and
past domains of dependence cover the entire spacetime; then data on S
determines the complete spacetime structure, and S is called a Cauchy
surface\index{Cauchy problem!Cauchy surface} for the spacetime.

Arnowitt, Deser, and Misner \cite{ArnDesMis62} developed the ADM
(Hamiltonian based) formalism\index{ADM formulation} showing precisely what initial data was needed
on S, and what constraint equations\index{constraint equations} it had to satisfy, in order that the
spacetime development from that data would be well defined. The initial
value problem\index{initial value problem} is to determine what initial data satisfies these constraints.
They also formulated the evolution equations needed to determine the time
development of this data. Lichnerowicz, Choquet-Bruhat, and Geroch showed
existence and uniqueness of a maximal Cauchy development can be proved using
functional analysis techniques based on Sobolev spaces. These existence and
uniqueness theorems are discussed in Chapter 8.

The study of asymptotically flat spacetimes leads to positive mass theorems.
Associated with these theorems are non-linear stability results for the
lowest energy solutions of Einstein's equations [the Minkowski and de
Sitter space-times] that are discussed in Chapter 9.

\subsection{Singularity theorems}
\index{singularity theorems|(}

Singularities - an edge to spacetime - occur in the Schwarzschild solution
and in the standard \index{FLRW models} FLRW models of
cosmology. A key issue is whether these are a result of the high symmetry of
these spacetimes, and so they might disappear in more realistic models of
this situations. Many attempts to prove theorems in this regard by direct
analysis of the field equations and examination of exact solutions failed.
The situation was totally transformed by a highly innovative paper by Roger
Penrose in 1965 \cite{Pen65} that used global methods and causal analysis to
prove that singularities will occur in gravitational collapse situations
where closed trapped surfaces occur, a causality condition is satisfied, and
suitable energy conditions are satisfied by the matter and fields present. A
closed trapped surface occurs when the gravitational field is so strong that
outgoing null rays from a 2-sphere converge - which occurs inside $r=2m$ in
the Schwarzschild solution. There are various positive energy
conditions\index{energy conditions}
that play a key role in these theorems, for example the null energy
condition $\rho +p\geq 0$, where $\rho $ is the energy density of the matter
and $p$ its pressure \cite{HawEll73}. Instead of characterising a
singularity by divergence of a scalar field such as the energy density, it
was characterized in this theorem by geodesic
incompleteness\index{geodesic incompleteness}: that is, some
timelike or null geodesics could not be extended to infinite affine
parameter values, showing there is an edge to spacetime that can be reached
in a finite time by freely moving articles or photons. Thus their possible
future or past is finite. Penrose's paper proved that the occurrence
of black hole
singularities is not due to special symmetries, but is generic.

This paper opened up entirely new methods of analysis and showed their
utility in key questions. Its methods were extended by Hawking, Geroch,
Misner, Tipler, and others; in particular Hawking proved similar theorems
for cosmology, in effect using the fact that time-reversed closed trapped
surfaces occur in realistic cosmological models; indeed their existence can
be shown to be a consequence of the existence of the cosmic microwave
blackbody radiation \cite{HawEll73}. A unifying singularity theorem was
proved by Hawking and Penrose \cite{HawPen70}. The nature of energy
conditions and the kinds of singularities that might exist were explored,
leading to characterization of various classes of singularities (scalar,
non-scalar, and locally regular), and alternative proposals for defining
singularities were given particularly by Schmidt.

\subsubsection{A `major crisis for physics'}

John Wheeler emphasized that existence of spacetime singularities - an edge
to spacetime, where not just space, time, and matter cease to exist, but
even the laws of physics themselves no longer apply - is a major crisis for
physics:

\begin{quote}
\textit{The existence of spacetime singularities represents an end to the
principle of sufficient causation and to so the predictability gained by
science. How could physics lead to a violation of itself -- to no physics? }%
\cite{CurBok12}
\end{quote}

This is of course a prediction of the classical theory. It is still not
known if quantum gravity solves this issue or not.
\index{singularity theorems|)}

\subsection{Conclusion}

Because GR 
moves spacetime from being a fixed
geometrical background arena within which physics takes place to being a
spacetime that is a dynamical participant in physics, and replaces Euclid's
Parallel Postulate by the geodesic deviation equation, it radically changes
our understanding of the nature of space-time geometry. Because it conceives
of gravity and inertia as being locally indistinguishable from each other,
it radically changes our view of the nature of the gravitational force. The
resulting theory has been tested to exquisite precision
\cite{Wil06,Wil79a}; see also Chapter 2..
Because the curvature of spacetime allows quite different global
properties than in flat spacetime, it is possible for closed timelike lines
to occur. Because it allows for a beginning and end to spacetime, where not
just matter but even spacetime and the laws of physics cease to exist, it
radically alters our views on the nature of existence. What it does not do
is give any account of how spacetime might have come into existence: that
is beyond its scope.


\section{The Schwarzschild solution and the solar system}
\index{Schwarzschild solution|(}%
\label{Schw} 
Karl Schwarzschild developed his vacuum solution of the Einstein equations
in late 1915, even before the General Theory of Relativity was fully
developed \cite{FroPoiCli14}. This is one of the most important solutions in
general relativity 
 because of its application to the Solar system (this
section), giving very accurate tests of general relativity, because of the
remarkable properties of its maximal analytic extension, and because of its
application to black hole theory (next sections).

\subsection{The Schwarzschild exterior solution}

Consider a vacuum, spherically symmetric solution of the Einstein Field
Equations. To model the exterior field of the sun in the solar system, or of
any static star, we look for a solution that is\newline

(1) spherically symmetric (here we ignore the rotation of the sun and its
consequent oblateness, leading to a slightly non-spherical exterior field,
as well as the small perturbations due to the gravitational fields of the
planets),

(2) vacuum outside some radius $r_{S}$ representing the surface of a central
star or other massive object. Thus we consider the \textit{exterior
solution: }$R_{bf}=0$, ignoring the gravitational field of dust particles,
the solar wind, planets, comets, etc. It can be attached at $r_{S}$ to a
corresponding interior solution for the star that generates the field, for
example the Schwarzschild interior solution \cite{MisThoWhe73} which has constant
density (but does not have a realistic equation of state).

We can choose coordinates for which the metric form is manifestly
spherically symmetric: 
\begin{equation}
ds^{2}=-A(r,t)dt^{2}+B(r,t)dr^{2}+r^{2}(d\theta ^{2}+\sin ^{2}\theta \,d\phi
^{2})  \label{eq:Sss}
\end{equation}%
where $x^{0}=t$, $x^{1}=r$, $x^{2}=\theta $, $x^{3}=\phi $ (the symmetry
group $SO(3)$ acts on the 2-spheres $\{r=const,t=const\}$). A major result
is the \textit{Jebsen-Birkhoff theorem}: the solution necessarily has an
extra symmetry (there is a further Killing vector) and so is necessarily
static in the exterior region: $A=A(r),B\ =B(r)$, independent of the time
coordinate $t$ \cite{MisThoWhe73,HawEll73}. This is a local result:\
provided the solution is spherically symmetric, it does not depend on
boundary conditions at infinity. It implies that spherical objects cannot
radiate away their mass (that is, there is no dilaton in general relativity
theory). Define $m\equiv MG/c^{2}>0$ (mass in geometrical units, giving the
one essential constant of the solution); then, setting $c=1$,

\begin{equation}
ds^{2}=-(1-{\frac{2m}{r}})dt^{2}+(1-{\frac{2m}{r}})^{-1}dr^{2}+r^{2}(d\theta
^{2}+\sin ^{2}\theta \,d\phi ^{2})  \label{Sch_metric}
\end{equation}%
This is the Schwarzschild (Exterior) solution \cite{Sch16}. It is an \textit{%
exact} solution of the EFE (no linearization was involved) for the exterior
field of a central massive object. It is valid for $r>r_{S}$ where $r_{S}$
is the coordinate radius of the surface of the object; we require that $%
r_{S}>r_{G}\equiv 2m\equiv 2MG/c^{2}$ where $r_{G}$ is the gravitational
radius or \textit{Schwarzschild radius} of the object; this is the mass in
geometrical units. If $r_{S}<r_{G}$, we would have a black hole and the
interior/exterior matching would be impossible.

It is \textit{asymptotically flat}: as $r\rightarrow \infty $, (\ref%
{Sch_metric}) becomes the metric of Minkowski spacetime in spherically
symmetric coordinates. Note that we did not have to put this in as an extra
condition: it automatically arose as a property of the exact solution.

\subsection{Effects on particles}

The spacetime geometry determines how particles move in the spacetime.

\subsubsection{Particle Orbits and Light rays}

The importance of this solution is that it determines the paths of particles
and light rays in the vicinity of the Sun. To determine them one needs to
solve the geodesic equations for timelike and lightlike curves. The
Lagrangian for geodesics is $L=g_{ab}(x^{c})\dot{x}^{a}\dot{x}^{b}$ where $%
\dot{x}^{a}=dx^{a}/dv$, with $v$ an affine parameter along the geodesics.
Applying this to timelike paths, one determines orbits for planets around
the Sun, obtaining standard bound and unbound planetary orbits but with
perihelion precession, as confirmed by observations \cite{MisThoWhe73}; applying
it to photon orbits, one gets equations both for gravitational redshift and
for gravitational lensing, again as confirmed by observations \cite{MisThoWhe73}.

Putting them together, one gets the predictions for laser ranging
experiments that have confirmed the predictions of general relativity theory
with exquisite precision \cite{Wil06}. One can also determine the proper
time measured along any world line and combine it with the gravitational
redshift and Doppler shift predictions for signals from a satellite to
Earth, thereby providing the basis for precision GPS systems \cite{Ash03}.
Thus GR plays a crucial role in the accuracy of useful GPS systems.

\subsubsection{Spin precession}

Parallel transport along a curve corresponds to Fermi-Walker transport along
the curve, leading to the prediction of spin precession, or frame dragging.
This can be measured by gyroscope experiments, as tested recently by gravity
probe B \cite{Wil11}. The effect is also observable in binary pulsars \cite%
{BreKasKra08}.

\subsection{Solar system tests of general relativity}

This set of results enables precision tests of General Relativity in the
Solar Systems. Clifford Will and others have set up a Parameterized
Post-Newtonian (PPN) formalism \cite{Wil06}. whereby general relativity can
be compared to other gravitational theories through this set of experiments,
that is 1. Planetary orbits; 2. Light bending; 3. Radar ranging; 4. Frame
Dragging experiment; 5. Validation of GPS systems. These solar system tests
of GR establish its validity on solar system scales to great accuracy. They
are discussed in detail in Chapter 2. 

Together these tests confirm that Einstein's gravitational theory - a
radical revolution in terms of viewing how gravitation works, developed by
pure thought from careful analysis of the implications of the equivalence
principle -- is verified to extremely high accuracy. There is no
 experimental evidence that it is wrong.
\index{Schwarzschild solution|)}

\subsection{Reissner-Nordstr\"om, Kottler, and Kerr Solutions}

There are three important generalizations of the Schwarzschild solution.

\subsubsection{Reissner-Nordstr\"om solution}%
\index{Reissner-Nordstr\"om solution|(}

Firstly, there is the Reissner-Nordstr\"om solution, which is the charged
version of the Schwarzschild solution \cite{Rei16,Nor18}, where in (\ref%
{eq:Sss}) the factor $A(r,t)=1-\frac{2m}{r}+\frac{GQ^{2}}{\varepsilon r^{2}}%
=1/B(r,t)$. It is spherically symmetric and static, but has a non-zero
electric field due to a charge $Q$ at the centre. It is of considerable
theoretical interest, but is not useful in astrophysics, as stars are not
charged (if they were, electromagnetism rather than gravitation would
dominate astronomy).
\index{Reissner-Nordstr\"om solution|)}

\subsubsection{Kottler solution}
\index{Kottler solution|(}

Secondly, the generalization of this spacetime to include a cosmological
constant $\Lambda $ is the Kottler solution \cite{Kot18}, where in (\ref%
{eq:Sss}) the factor $A(r,t)=1-\frac{2m}{r}-\frac{\Lambda r^{2}}{3}=1/B(r,t)$%
, which is a Schwarzschild solution with cosmological constant. It gives the
field of a spherical body imbedded in a de Sitter universe. This is relevant
to the present day universe, because astronomical observations have detected
the effect of an effective cosmological constant in cosmology at recent
times.
\index{Kottler solution|)}

\subsubsection{Kerr solution}\index{Kerr solution|(}

Finally there is the Kerr solution \cite{Ker63,KerSch65} which is
the rotating version of the Schwarzschild solution. It is a vacuum solution
that is stationary rather than static and axially symmetric rather than
spherically symmetric \cite{MisThoWhe73,HawEll73}. It is of considerable
importance because most astrophysical objects are rotating. There is one
important difference from Schwarzschild: while we can construct exact
interior solutions to match the Schwarszchild exterior solution, that is not
the case for the Kerr solution. It has a complex and fascinating structure
that is still giving new insights \cite{AbdLak13}.

The geodesic structure of these spaces can be examined as in the case of the
Schwarzschild solution. Carter showed that in the case of the Kerr solution,
there were hidden integrals for geodesics associated with existence of
Killing tensors \cite{Car68a,Car73}.
\index{Kerr solution|)}


\subsection{Maximal Extension of the Schwarzschild solution}
\index{Schwarzschild solution!maximal extension|(}

\label{max_ext} 
The previous section considered the solution exterior to the surface of a
star, and this was all known and understood early on, certainly by the mid
1930s. The behaviour at the coordinate singularity at $r = 2m$ is another
matter: despite useful contributions by Eddington, Finkelstein, Szekeres,
and Synge, it was only fully understood in the late 1960s through papers
by Kruskal and Fronsdal (see \cite{Eis82,Eis87} for the early historical development).

Consider now the Schwarzschild solution as a vacuum solution with no central
star. It is then apparently singular at $r=2m$, so has to be restricted to $%
r>2m$; then both ingoing and outgoing null geodesics are incomplete, because
they cannot be extended beyond $r=2m$. However the scalar invariant $%
R_{abcd}\,R^{abcd}~=~{\frac{48m^{2}}{r^{6}}}$ is finite there, and it turned
out this is just a coordinate singularity: one can find regular coordinates
that extend the solution across this surface, which is in fact a null
surface where the solution changes from being static but spatially
inhomogeneous (for $r>2m$) to being spatially homogeneous but time varying
(for $0\leq $ $r<2m).$

Eddington, Szekeres, Finkelstein, and Novikov showed how one could find
regular coordinates that cross the surface $r=2m.$ To do so, one can note
that radial null geodesics are given by ${\frac{dt}{dr}}=\pm \,{\frac{1}{1-{%
\frac{2m}{r}}}}~\Leftrightarrow ~t=\pm ~r^{\ast }+const$ where $dr^{\ast }={%
\frac{dr}{1-{\frac{2m}{r}}}}~\Leftrightarrow ~r^{\ast }=r+2m\,\ln \left( {%
\frac{r}{2m}}-1\right) .$ Defining $v=t+r^{\ast },~~w=t-r^{\ast },$ one can
obtain three null forms for the Schwarzschild metric. Changing to
coordinates $(v,r,\theta ,\phi )$, the metric is 
\begin{equation}
ds^{2}=-(1-{\frac{2m}{r}})dv^{2}+2dvdr~+~r^{2}d\Omega ^{2}
\end{equation}%
which is the Eddington-Finkelstein form of the metric \cite{Edd24}. The
coordinate transformation has succeeded in getting rid of the singularity at 
$r=2m$, and shows how infalling null geodesics can cross from $r>2m$ to $%
r<\ 2m$, but the converse is not possible, as can be seen in the
Eddington-Finkelstein diagram (Figure 1), which shows how the light
cones tip over at
different radii. This is thus a \textit{black hole} because light cannot
escape from the interior region $r<2m$ to the exterior region $r>2m.$ The
coordinate surfaces $\{r=const\}$ are timelike for $r>2m$, null for $r=2m,$
and spacelike for $r<2m$; the coordinate surfaces $\{t=const\}$ are
spacelike for $r>2m$, undefined for $r=2m,$ and timelike for $r<2m$. This
warns us not to expect that a coordinate $t$ will necessarily be a good time
coordinate everywhere just because we call it "time".

\begin{center}
\includegraphics[width=4.0in]{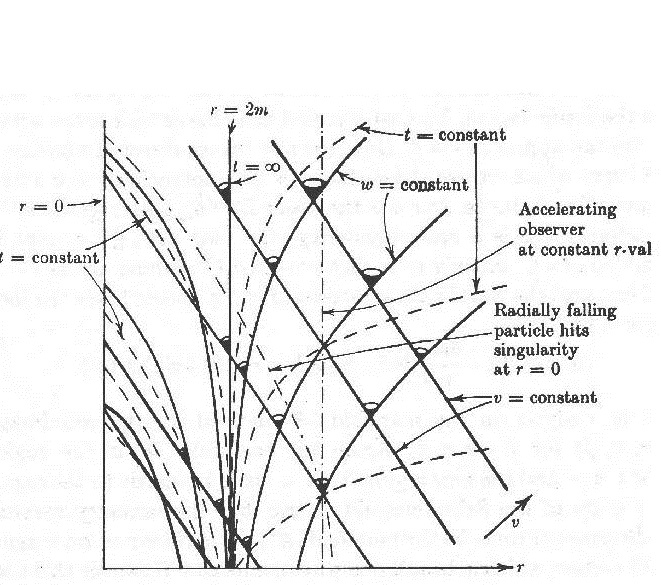}
\end{center}
\textbf{Figure 1}. Eddington-Finkelstein diagram of the Schwarzschild solution. Local light cones are shown.\\

The
coordinate transformation (which is singular at $r=2m$) extends the original
space-time region \textbf{I}, defined by $2m<r<\infty $, to a new spatially
homogeneous but time dependent region \textbf{II}, defined by $0<r<2m.$
Ingoing null geodesics are then complete; but outgoing ones are not complete
in the past. Use of coordinates $(w,r,\theta ,\phi )$ by contrast adds a new
region \textbf{II' }that makes the outgoing geodesics complete but
the ingoing ones complete. Using coordinates $(v,w,\theta ,\phi )$ and then
conformally rescaling these coordinates makes both sets of null geodesics
complete but to make the geodesics in regions \textbf{II} and \textbf{II'}
complete one must add a further region \textbf{I' }that completes the past
directed null geodesics of \textbf{II} and the future directed null
geodesics of \textbf{II'}.

Thus this shows that the maximally extended solution \cite{Kru60} consists
of these four domains separated by null surfaces $r=2m$ that form the event
horizons of the black hole, which has a non-trivial topology: two
asymptotically flat regions back to back, connected by \textquotedblleft
wormholes\textquotedblright\ \cite{MisThoWhe73}; it is maximally extended because
all geodesics either extend to infinity, or end on a past or future
singularity (Figure 2.). Taking a cross section $t^{\prime
}=const$,
for
large negative values of $t^{\prime }$ one has two separated asymptotically
flat regions \textbf{I}, \textbf{I}'; for $-2m<\ t^{\prime }<\ 2m$, the two
asymptotically flat regions are connected by a throat; for $t^{\prime }>\ 2m$
they are again separated from each other. This opening and closing bridge
between them is a wormhole joining the two asymptotically flat regions
\index{Penrose diagram!Schwarzschild solution}\index{Kruskal diagram}.

\begin{center}
\includegraphics[scale=0.6,keepaspectratio]{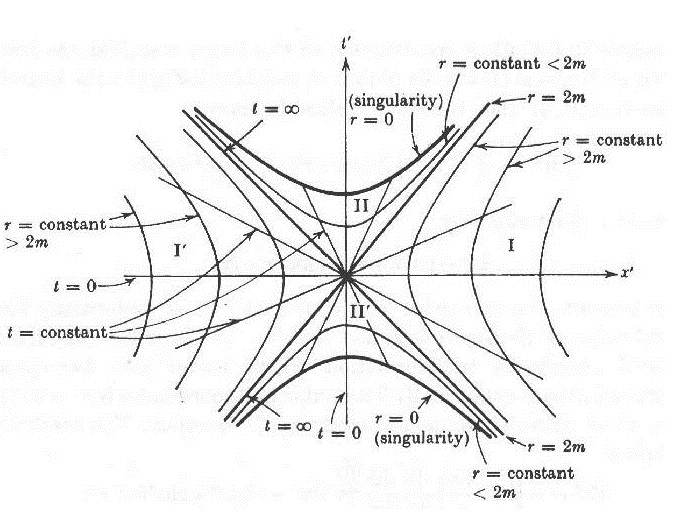}
\end{center}
Figure 2. The Kruskal maximal extension of the Schwarzschild vacuum solution.\\

The
singularities are spacelike boundaries to spacetime (one in the future and
one in the past), not timelike world lines as one would expect. The
coordinate singularities in the original Schwarzschild form of the metric
are because the time coordinate $t$ diverges there. Use of conformal
transformations of the null coordinates yields the Penrose diagram of the
maximally extended spacetime including its boundaries at infinity \cite%
{HawEll73,MisThoWhe73}.

\subsubsection{Symmetries}

The solution is invariant under

(a) A left-right symmetry, where regions I and I' are identical to each
other, while regions II and II' are individually symmetric under the
interchange $t^{\prime }\rightarrow $ $-t^{\prime };$

(b) A time symmetry, where regions II and II' are identical to each other,
while regions I and I' are individually symmetric under the interchange $%
x^{\prime }\rightarrow $ $-x^{\prime };$

(c) The boost symmetry that is shown to exist by Birkhoff's theorem, with
timelike orbits in regions I and I', spacelike orbits in regions II and II',
and null orbits in the four null horizons (Killing vector orbits) that
bifurcate at the central 2-sphere, which is a set of fixed points of the
group. This saddle-point behaviour is a generic property of bifurcate
Killing horizons, where the affine parameter and Killing vector parameter
are exponentially related to each other \cite{Boy69}.
\index{Schwarzschild solution!maximal extension|)}

\subsubsection{Reissner-Nordstr\"om, Kottler, and Kerr Solutions}
\index{Reissner-Nordstr\"om solution}%
\index{Kottler solution}\index{Kerr solution}

As shown by Carter \cite{Car68} and Boyer and Lindquist \cite{BoyLin67}, one
can obtain similar maximal extensions of the Reissner-Nordstr\"om and Kerr
solutions \cite{Car68,HawEll73}. They are far more complex than
those for the Schwarzschild solution, having many horizons and
asymptotically flat regions. Lake and Roeder have given the maximal
extension for the Kottler spacetime \cite{LakRoe77}.

\subsection{Conclusion}

The Schwarzschild solution\index{Schwarzschild solution}
 enables us to model the geometry of the solar
system with exquisite precision, giving a more accurate description of solar
system dynamics than Newtonian theory does. This enables us to test the
static aspects of general relativity to very high accuracy, and so confirm
its correctness as one of the fundamental theories of physics.

The maximally extended Schwarszchild solution is an extraordinary discovery.
The very simple looking metric (\ref{Sch_metric}) implies the  existence of
two asymptotically flat spacetime regions connected by a wormhole; event
horizons (the null surfaces $r=2m$) separating the interior ($r<\ 2m$) and
exterior ($r\ >\ 2m$) regions; and two singularities that are spatially
homogeneous in the limit $\ r\rightarrow 0.$ It is impossible for a
maximally extended solution to be static. There is no central worldline, as a
point particle picture suggests. Thus just as quantum physics implied a
radical revision of the idea of a particle, so does general relativity:
there is no general relativity version of the Newtonian idea of a point
particle.

None of this is obvious. The global topology is not optional; it follows
from the way the Einstein equations for this vacuum curve spacetime. And the
nature of this solution emphasizes why one should always try to determine
exact properties of solutions in general relativity: the global properties
of the linearised form of the Schwarzschild solution (which does not exactly
satisfy the field equations) will be radically different.


\section{Gravitational collapse and black holes}\index{gravitational collapse}

\label{BH} 
Astrophysical black holes exists because they arise by the collapse of
matter due to its gravitational self-attraction.

\subsection{Spherically symmetric matter models}

Spherically symmetric fluid filled models enable one to investigate how
gravitational attraction makes matter aggregate from small perturbations
into major in$\hom $ogeneities that enter the non-linear regime and form
black holes. Pressure-free models one can use to investigate this are the
Lema\^{\i}tre-Tolman-Bondi models \cite{Tol34,Bon47,Kra06}
which evolve inhomogeneously with spherical shells each obeying a version of
the Friedmann equation of cosmology. One can use these as interior
solutions, with the Schwarzschild solution as the exterior (vacuum)
solution, joined across a timelike surface where the standard junction
conditions (Section \ref{sec:join}) are fulfilled. More realistic models
with pressure obey the Oppenheimer-Volkov equation of evolution \cite{MisThoWhe73}%
. Astrophysical studies show that if the collapsing object is massive enough
(its mass is greater than the Chandrasekhar limit), there is no physical
pressure that will halt the collapse: the final state will be a black hole 
\cite{MisThoWhe73,BegRee10}. Using pressure-free models to investigate
gravitational collapse, Oppenheimer and Snyder showed this would eventually
occur: "\textit{we see that for a fixed value of }$R$ \textit{as }$t$%
\textit{\ tends toward infinity, }$\tau $\textit{\ tends to a finite limit,
which increases with }$R$\textit{. After this time }$\tau _{0}$\textit{\ an
observer comoving with the matter would not be able to send a light signal
from the star; the cone within which a signal can escape has closed entirely}%
." \cite{OppSny39}.

\subsection{Causal diagrams}
\index{Penrose diagram!collapsing fluid spheres}

Basically the situation is like Figure 1, but the solution's central part of
the diagram is no longer vacuum; it is cut off by the infalling fluid.
Initially the surface of the fluid is at $r_{S}>2m$ so there are no closed
trapped surfaces. Then the fluid surface crosses the value $r_{S}=2m$ and
thereafter lies inside the event horizon. Light emitted from the surface of
the star at later times is trapped behind the event horizon and cannot
emerge to the outside world.

The space-time diagram showing how this occurs is given in Figure
3: the
event horizon at $r=2m$ is a null surface, so light emitted at that radius
never moves inwards or outwards; the gravitational attraction of the mass at
the centre holds it at a constant distance from the centre. From the
outside, the collapse never seems to end; there is always light arriving at
infinity from just outside the event horizon, albeit with ever increasing
redshift and hence ever decreasing intensity.

\begin{center}
\includegraphics[scale=1.0,keepaspectratio]{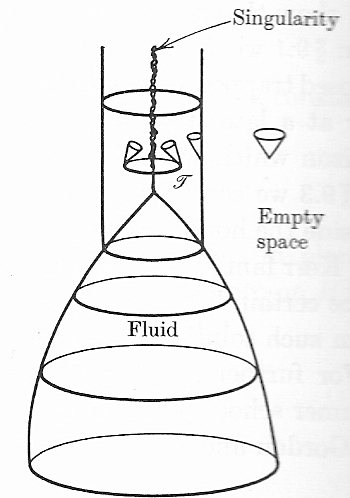}
\end{center}
Figure 3. Collapse to a black hole.\\

However this diagram is misleading in some ways: it suggests that
the central singularity is a timelike world line, which is not the case; it
is spacelike because it exists in the part of spacetime corresponding to
region II in Figure 2. The Penrose diagram for what happens is shown in
Figure 4; it shows that the outer regions are the
same as region I in Figure 2,
but existence of the infalling star leads to a regular centre rather than
another asymptotically flat region $I^{\prime }$. A key feature, pointed out
by Penrose, is that closed trapped surfaces exist for region II given by $%
r<2m$: the area spanned by outgoing null geodesics from each 2-sphere
decreases as one goes to the future. This is the reason that the occurrence
of a singularity in the future is inevitable \cite{Pen65,CurBok12}.

 \begin{center}
\includegraphics[scale=1.0,keepaspectratio]{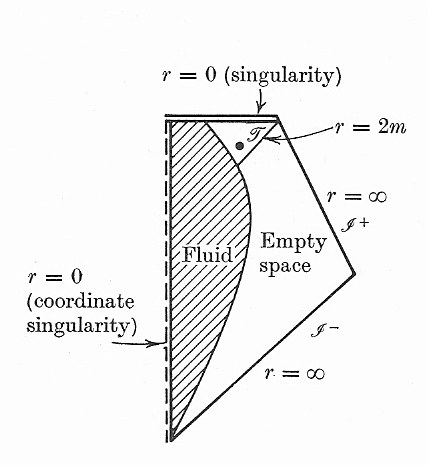}
 \end{center}
Figure 4. Penrose diagram
	of collapse to a black hole.\\

The question
then is, how generic is this situation where the final singularity is hidden
behind the horizon and so is invisible to the external world: does this
only occur in spherically symmetric spacetimes? A very innovative uniqueness
theorem by Israel \cite{Isr68} stated that black holes must necessarily be
spherically symmetric if a regular event horizon forms. But then the
question is, how general is formation of an event horizon? Penrose
formulated the cosmic censorship hypothesis, that such horizons would indeed
form in the generic case \cite{Pen99a}. This conjecture is still unresolved;
it is discussed in Chapter 9.

Much effort has been extended in showing that the Kerr solution is the
likely final state of gravitational collapse of a rotating object. Work by
Hawking, Carter, Robinson and others shows this indeed seems to be the case
\cite{Rob75}. Black hole uniqueness theorems are discussed in chapter 9.

An important feature in such collapse is the Penrose inequality relating
mass and black hole horizon areas, discussed in Chapter 8, and Hawking's
area theorem\index{area theorem}, which states that the area of cross sections of a black hole
horizon is non-decreasing towards the future. This leads to the proposed
laws of black hole thermodynamics, associating a temperature and entropy to
black holes in a way that is parallel to the usual laws of thermodynamics 
\cite{BarCarHaw73}, in particular the temperature is $T_{H}=\frac{\kappa }{%
2\pi }\ $where $\kappa$ is the surface gravity of the hole, and the black
hole entropy is proportional to the area of its event horizon divided by the
Planck area: $S_{BH}=\frac{kA}{4\pi l_{P}^{2}}$ where $l_{P}$ is the Planck
length$.$ This led on to Hawking's famous discovery of the emission of black
body radiation by black holes through quantum field theory processes; that
development however lies outside the scope of this chapter.

\subsection{Accretion discs and Domains}

Black holes\index{black hole} will generally be surrounded by accretion discs%
\index{accretion!disk} that will emit
X-rays due to viscous heating as the matter falls in towards the event
horizon (see Chapter \ref{ch:relastro}. Rotating black holes have
three important domains affecting this process \cite{AbrFra13}:

1. \textit{The Event Horizon}: That radius inside of which escape from the
black hole is not possible;\index{event horizon}

2. T\textit{he Ergosphere}: That radius inside of which negative energy
states are possible (giving rise to the potentiality of tapping the energy
of the black hole).\index{ergo region}

3.\textit{ Innermost Stable Circular Orbit} (ISCO): That radius inside of
which free circular orbital motion is not
possible.\index{ISCO}\index{innermost stable circular orbit|see{ISCO}}

These are the geometric features on which the theory of accretion discs is
based.

\subsection{Powerhouses in astrophysics}

Black holes are intriguing aspects of general relativity, and theoretical
studies showed they were indeed likely to form in astrophysical situations.
But are they relevant to the universe out there?

It seems indeed so. They were discovered theoretically as an
unexpected consequence of the maximal extensions of the Schwarzschild
solution. They became central to high energy astrophysics, as shown by
Lynden Bell and Rees, being the key to understanding of
QSOs \cite{Ree84,BegRee10}.  Furthermore supermassive black
holes\index{black hole!supermassive} occur in many galaxies,
surrounded by accretion disks \cite{FerMer02}, while stellar mass
black holes\index{black hole!stellar} occur as the endpoint of the
lives of massive stars \cite{Pen96}. There is much astronomical
evidence for their existence \cite{CelMilSci99} including one at the
centre of our own galaxy \cite{SchOttGen02,Mel07}. The accretion disks
that emit radiation whereby we can detect them can be modelled in a
general relativity way \cite{AbrFra13}. This is all discussed in
Chapter 3 on relativistic astrophysics, considering
the observational status of black holes, and their relation to gamma
ray bursts.  Chapter 7 discusses how numerical
simulations\index{numerical simulations} have provided insights into
the critical phenomenon at the threshold of black hole formation,
gravitational collapse, mergers of black holes, and mergers of black
holes and neutron stars.

\subsection{Conclusion}

Black holes were theoretically predicted to occur because they are solutions
of the Einstein Field Equations, but have turned out to play a key role in
high energy astrophysics \cite{BegRee10}. They occur as the endpoint of
evolution of massive stars, and occur at the centre of galaxies, where they
play a key role in galaxy dynamics and provide the powerhouse for high
energy astrophysical phenomena that are otherwise inexplicable. Furthermore,
black hole interactions are expected to provide the source of gravitational
waves that will enable us to probe the extremely early universe with great
precision (Section \ref{sec:gravwavecosm}). They have no analogue in
Newtonian gravitational theory because there is no limit to the speed of
propagation of signals in that theory (where the speed of light plays no
special role).


\section{Cosmology and structure formation}

\label{FLRW} 
The development of cosmological models based in General Relativity theory
provides the basis for our understandings of both the evolution of the
universe as a whole, and of structure formation\index{structure formation} within it. The start was
Einstein's static universe model of 1917 \cite{Ein17}, followed by the
static de Sitter universe in 1917, Friedmann's expanding models of 1922 \cite%
{Fri22} and 1924 \cite{Fri24}, and Lemaitre's expanding model of 1927 \cite%
{Lem31,Lem33}. These expanding models were ignored until Eddington's
proof of the instability of the Einstein static universe in 1930 \cite{Ell90a}%
. The idea of the expanding universe was then generally accepted, and
canonised in Robertson's fine review article in 1933 \cite{Rob33} (which
gives annotated references to all the earlier papers). It is noteworthy that
the first Newtonian cosmological models were developed only in 1935 by Milne
and McCrea - some 18 years after the first general relativistic models.

Today the `standard model of cosmology', based in the way general relativity
shows how matter curves spacetime, is highly successful in describing
precision observations of the large scale properties of the universe,
although some significant problems concerning the nature of the matter and
energy fields controlling the dynamics of the universe remain unresolved 
\cite{Dod03,PetUza13,EllMaaMac12}.

\subsection{FLRW models and the Hot Big Bang}
\index{FLRW models|(}%

The expanding universes of Friedmann and Lemaitre (`FL') are a family of
exact solutions of the Einstein field equations that are spatially
homogenous and isotropic. The geometry of these standard models was
clarified by Robertson and Walker (`RW') in 1935. The metric is
characterised by a scale factor $a(t)$ representing the time change of the
relative size of the universe:

\[
ds^{2}=-dt^{2}+a^{2}(t)d\sigma ^{2},\;d\sigma ^{2}=dr^{2}+f^{2}(r)d\Omega
^{2} 
\]%
with $f(r)=$ $\left\{ \sin r,r,\sinh r\right\} $ if $k=\left\{
+1,0,-1\right\} $ and $d\Omega $ $^{2}=d\theta ^{2}+\sin ^{2}\theta d\phi
^{2}$; thus the 3-space metric $d\sigma ^{2}$ represents a 3-space of
constant curvature $k.$ Spatial homogeneity together with isotropy implies
that a multiply transitive group of isometries $G_{6}$ acts on the surfaces $%
\left\{ t=constant\right\} ,$ and consequently all physical and geometric
quantities depend only on $t.$ The 4-velocity $u^{a}=dx^{a}/dt$ of preferred
fundamental observers is 
\[
u^{a}=\delta _{0}^{a}\Rightarrow u^{a}u_{a}=-1 
\]%
so the time parameter $t$ represents proper time measured along fundamental
world lines $x^{a}(t)$ with $u^{a}$ as tangent vector; the distances between
these world lines scales with $a(t)$, so volumes scale as $a^{3}(t)$. \ \
The relative expansion rate of matter is represented by the Hubble parameter 
$H(t)=(1/a)(da/dt),$ and the rate of slowing down by the deceleration
parameter $q:=-(1/a)(d^{2}a/dt^{2})$. These space times are conformally
flat: $C_{abcd}=0$ (there is no free gravitational field, so no tidal
forces or gravitational waves occur in these models).

\subsubsection{Dynamics}

The behaviour of matter and expansion of the universe are governed by three
related equations. First, the energy conservation equation (\ref{eq:cons})
becomes 
\begin{equation}
d\rho /dt+3H(\rho +p)=0  \label{cons}
\end{equation}%
relating the rate of change of the energy density $\rho (t)$ to the pressure 
$p(t).$It becomes determinate when we are given an equation of state $%
p=p(\rho ,\phi )$ determining $p$ in terms of $\rho $ and possibly some
internal variable $\phi $ (which if present, will have to have its own
dynamical equations). The normalized density parameter is $\Omega :=\frac{%
\kappa \rho }{3H^{2}}.$ The Raychaudhuri equation (\ref{ray2}) becomes

\begin{equation}
\frac{3}{a}\frac{d^{2}a}{dt^{2}}=-{\textstyle{\frac{\kappa }{2}}}(\rho
+3p)+\Lambda  \label{Ray}
\end{equation}%
which directly gives the deceleration due to matter, and shows (i) the
active gravitational mass density is $\rho _{grav}:=\rho +3p$, which is
positive for all ordinary matter, and (ii) a cosmological constant $\Lambda 
$ causes acceleration iff $\Lambda >0.$ Its first integral is the Friedmann
equation 
\begin{equation}
3H^{2}=\kappa \rho +\Lambda -{\textstyle{\frac{3k}{a^{2}}}},  \label{Fried}
\end{equation}%
where ${{\frac{3k}{a^{2}}}}$ is the curvature of the 3-spaces; this is just
the Gauss-Codazzi equation relating the curvature of imbedded 3-spaces to
the curvature of the imbedding 4-dimensional spacetime \cite{Ehl61,Ell71}. The evolution of $a(t)$ is determined by any two of (\ref{cons}), (%
\ref{Ray}), and (\ref{Fried}).

The basic behaviour has been known since the 1930s \cite{Rob33}. Normal
matter can be represented by the equation of state $p=(\gamma -1)\rho $,
where $\gamma =1$ for pressure free matter (`dust' or `baryonic matter'); $%
\gamma =4/3$ for radiation; and $\gamma =-1$ is equivalent to a cosmological
constant. There is always a singular start to the universe at time $%
t_{0}<1/H_{0}$ ago if $\rho +3p>0$, which will be true for ordinary matter
plus radiation, and $\Lambda \leq 0$; however one can get a bounce if both $%
\Lambda >0$ and $k=+1$ (for matter with $\rho >0,$ bounces can only occur if 
$k=+1)$. When $\Lambda =0,$ if $k=+1$, $\Omega >1$ the universe recollapses;
if $k=-1$, $\Omega <\ 1$ it expands forever; and $k=0$ ($\Omega =1)$ is the
critical case separating these behaviours, that just succeeds in expanding
forever. When $\Lambda >0$ and $k=+1$, a static solution is possible (the
Einstein static universe \cite{Ein18}), and the universe can bounce or
`hover' close to a constant radius (these are the Eddington-Lema\^{\i}tre
models). All these behaviours can be illuminatingly represented by
appropriate dynamical systems phase planes  \cite{EhlRin89,WaiEll97}%
.

\subsubsection{\textbf{Observational relations}}

Observational relations can be worked out for these models, based on the
fact that photons move on null geodesics $x^{a}(v)$ with tangent vector $%
k^{a}(v)\ =dx^{a}/dv:k^{a}k_{a}=0,$ $k_{\;;b}^{a}k^{b}=0$ \cite{KriSac66}.
This shows that the radial coordinate value of radial null geodesics through
the origin (which are generic null geodesics, because of the spacetime
symmetry) is given by $\{ds^{2}=0,\;d\theta =0,\;d\phi =0\}$ which gives
\begin{eqnarray}
u(t_{0},t_{1}):=\int_{0}^{r_{emit}}dr=\int_{t_{emit}}^{t_{obs}}\frac{dt}{a(t)}=\int_{a_{emit}}^{a_{obs}}\frac{1}{H(t)}\frac{da}{a^{2}(t)}. 
\end{eqnarray}
Substitution from the Friedmann equation (\ref{Fried}) shows how the
cosmological dynamics affects $u(t_{0},t_{1}).$ Key variables resulting are
observed redshifts $z,$ given by

\[
1+z=\frac{\left( k^{a}u_{a}\right) _{emit}}{\left( k^{b}u_{b}\right) _{obs}}=%
\frac{a(t_{obs})}{a(t_{emit})}, 
\]%
and area distance $r_{0},$ which up to a redshift factor is the same as the
luminosity distance $D_{L}:$ 
\[
D_{L}=r_{0}(1+z); 
\]%
this is the reciprocity theorem \cite{Eth33,Ell71}. One can work out
observational relations for galaxy number counts versus magnitude $(n,m)$
and the magnitude--redshift relation $(m,z)$, which determines the
present deceleration parameter $q_{0}$ when applied to \textquotedblleft
standard candles\textquotedblright\ \cite{San61}. A power series derivation
of observational relations in generic cosmological models is given by
Kristian and Sachs \cite{KriSac66}.

Major observational programmes have examined these relations and determined $%
H_{0}$ and $q_{0},$ showing that, assuming the geometry is indeed that of a
Robertson-Walker spacetime, the universe is accelerating at recent times ($%
q_{0}<0)$. This means some kind of dark energy is present such that at
recent times, $\rho +3p<\ 0$ \cite{Dod03,PetUza13,EllMaaMac12}. The simplest interpretation is that this is due to a
cosmological constant $\Lambda >0$ that dominates the recent dynamics of the
universe.

A key finding is the existence of limits both to causation, represented by
particle horizons, and to observations, represented by visual horizons. The
issue is whether $u(t_{0},t_{1})$ converges or diverges as $t_{0}\rightarrow
0;.$and for ordinary matter and radiation, it converges, representing a
limit to how far causal effects can propagate since the start of the
universe. Much confusion about their nature was cleared up by Rindler in a
classic paper \cite{Rin56}, with further clarity coming from use of Penrose
causal diagrams for these models \cite{Pen64}. This showed that particle
horizons would occur if and only if the initial singularity was spacelike.
There are many statements in the literature that such horizons represent
motion of galaxies away from us at the speed of light, but that is not the
case; they occur due to the integrated behaviour of light from the start of
the universe to the present day \cite{EllRot93}, with the visual horizon,
determined by the most distant matter we can detect by electromagnetic
radiation, lying inside the particle horizon. This is why the visual horizon
size can be 42 billion light years in an Einstein de Sitter model with a
Hubble scale of 14 billion years. Event horizons relate to the ultimate
limits of causation in the future universe, that is whether $u(t_{0},t_{1})$
converges or diverges as $t_{1}\rightarrow \infty $; while they play a key
role as regards the nature of black holes (Section \ref{max_ext}), they are
irrelevant to observational cosmology.

\subsubsection{Cosmological physics}

Cosmological models started off as purely geometrical, but then a major
realisation was that standard physics could be applied to the properties of
matter in the early universe.

First was the application of atomic physics to the expanding universe,
resulting in prediction of a Hot Big Bang (`HBB') early phase of the
universe with ionised matter and radiation in equilibrium with each other
because of tight coupling between electrons and radiation. This phase ended
when the temperature dropped below the ionisation temperature of the matter,
resulting in cosmic blackbody radiation being emitted at the Last Scattering
Surface at about $T_{emit}=4000K$. A major theoretical result, consequent on
the reciprocity theorem, is that the blackbody spectrum will be preserved
but with temperature $T_{obs}=T_{emit}/(1+z)$ \cite{Ell71}; hence this
radiation is observed today as cosmic microwave blackbody radiation (CMB)\
with a temperature of $2.73K$. There then follows a complex interaction of
matter and radiation in the expanding universe \cite{WagFowHoy67,%
SunZel70a}. A key feature is the evolution of the speed of sound with time,
as well as diffusion effects leading to damping of fine-scale structure \cite%
{Lon13}.

Second was the application of nuclear physics to the epoch of
nucleosynthesis\index{big bang nucleosynthesis} at a temperature of about $T=10^{8}$K, leading to
predictions of light element abundances resulting from nuclear interactions
in the early universe \cite{DorZelNov67,Pee66,WagFowHoy67}. The key point here
is that the Friedmann equation for early times, when radiation dominates the
dynamics and curvature is negligible, gives the temperature-time relation 
\cite{Beretal12}

\begin{equation}
T=\frac{T_{0}}{t^{1/2}},T_{0}:=0.74\left( \frac{10.75}{g_{\ast }}\right)
^{1/2}  \label{eq:tempcb}
\end{equation}%
with no free parameters. Here $t$ is time in seconds, $T$ is the temperature
in MeV, and $g_{\ast }$ is the effective number of particle species, which
is 10.75 for the standard model of particle physics, where there are
contributions of 2 from photons, 7/2 from electron-positron pairs and 7/4
from each neutrino flavor. It is this relation that determines the course of
the nuclear reactions, leading to formation of light elements (up to
Lithium) in the early universe. These element abundances agree with those
determined by astronomical observations, up to some unresolved worries about
Lithium. It was this theory that established cosmology as a solid branch of
physics \cite{Beretal12}. 

\subsubsection{In summary}

These models are the opposite of the Schwarzschild vacuum solution. Those
models represent the dynamics of pure vacuum (there is no matter tensor);
these models represent the dynamics of spacetime governed purely by matter
(there is no free gravitational field). Hence the outcome depends
irrevocably on the type of matter present, as shown for example in the
dependence (\ref{eq:tempcb}) of $T(t)$ on $g_{\ast }$. They are remarkably
simple, and remarkably good models of the real universe we observe by means
of astronomical observations. They also represent a key conundrum: they
predict a start to the universe, and indeed all of physics, at a spacetime
singularity. One key issue is whether that conclusion can somehow be avoided.
\index{FLRW models|)}%

\subsection{More general dynamics: Inflation}\index{inflation|(}

The previous section was based in well established physics. At earlier times
more speculative physics will necessarily be involved, involving
interactions not yet testable by particle accelerators.

More general matter dynamics will lead to more general behaviour of the
cosmological model at early times. In particular a scalar
field\index{scalar field} $\phi (t)$
with potential $V(\phi )$ may be present, obeying the Klein Gordon Equation
\begin{equation}
d^{2}\phi /dt^{2}+3H\phi +dV/d\phi =0.  \label{EllisKG} 
\end{equation}%
It will have an energy density $\rho =\frac{1}{2}(d\phi /dt)^{2}+V(\phi )$
and pressure $p=\frac{1}{2}(d\phi /dt)^{2}-V(\phi ),$ so the inertial and
gravitational energy densities are
\begin{equation}
\rho +p=(d\phi /dt)^{2}>0,\;\varrho +3p=(d\phi /dt)^{2}-V(\phi ).
\end{equation}%
Hence the active gravitational mass can be negative in the `slow roll' case: 
$(d\phi /dt)^{2}<V(\phi ).$ So scalar fields can cause an exponential
acceleration when $\phi $ stays at a constant value $\phi _{0}$ because of
the friction term $3H\phi $ in (\ref{EllisKG})  resulting in $\varrho +3p\approx
-V(\phi _{0})=const$. This is the physical basis of the \textit{inflationary
universe} idea of an extremely rapid exponential expansion at very early
times that smooths out and flattens the universe, also causing any matter or
radiation content to die away towards zero. The inflaton field itself dies
away at the end of inflation when slow rolling comes to an end and the
inflaton gets converted to matter and radiation by a process of reheating.%

It is now broadly agreed that there was indeed such a period of inflation in
the very early universe but the details are not clear: there are over 100
different variants \cite{MarRinVen13}, including single-field inflation,
multiple-field inflation, and models where matter is not described by a
scalar field as, for example vector inflation. As the potential is not tied
in to any specific physical field, one can run the field equations backwards
to determine the effective inflaton potential from the desired dynamic
behaviour \cite{EllMad91,LidLidKol97}. 
\index{inflation|)}

\subsection{Perturbed FLRW models and the growth of structure}
\index{FLRW models!perturbations|(}

The real universe is only approximately a Robertson-Walker spacetime.
Structure formation in an expanding universe can be studied by using
linearly perturbed FLRW models  at early times, plus numerical simulations
at later times when the inhomogeneities have gone
non-linear \cite{PeeYu70,Pee80}. The development of the theory of the
present structure 
formation, a major theoretical achievement, is described illuminatingly by
Longair (\cite{Lon13}, Chapter 15), which gives full references: it is only
possible to highlight a few of these papers in this brief section (and see
also Chapter 4 in this volume). 

Lema\^{\i}tre initiated the use of inhomogeneous models to study structure
formation \cite{Lem33a,Lem49,Lem58}. The study of general
linear perturbations of an expanding universe was initiated by Lifshitz \cite%
{Lif46}, and the path-breaking paper by Sachs and Wolfe \cite{SacWol67}
first studied their effect on microwave background radiation anisotropies.
That paper was vary careful about gauge freedom. However many others were
not, being plagued by the gauge problem: because of absence of a fixed
background spacetime, one can encounter unphysical gauge modes that are
essentially just coordinate fluctuations. A path-breaking paper by Bardeen
\cite{Bar80} developed a gauge covariant approach to linear perturbations
that enabled distinguishing of physical from coordinate modes, and this
approach \cite{MukFelBra92,KodSas84} is widely used. An alternative
1+3 covariant approach was initiated by Hawking \cite{Haw66} and then
developed by Ellis, Bruni, and collaborators \cite{EllBru89,BruDunEll92}. Both methods have been used to study both structure formation
and associated CMB anisotropies \cite{Dod03,Muk05,EllMaaMac12,PetUza13}. The perturbation equations can for example
be used to determine the dynamical effects of the cosmological constant \cite%
{LahLilPri91}. 

Key aspects of CMB anisotropies are the baryon-acoustic oscillations before
decoupling \cite{Sak65,SunZel70,PeeYu70}, which lead to the
observed CMB power spectrum peaks, and the Sunyaev-Zeldovich effect
(scattering of the CMB by ionised hot gas in clusters of galaxies) after
decoupling \cite{SunZel70a,SunZel80}, which acts as sensitive test
of conditions at the time of scattering. The perfection of the calculations
by Silk and Wilson \cite{SilWil79} and Bond and Efstathiou \cite{BonEfs84},
and the subsequent explosion of precision calculations (as opposed to
measurements) in the 90s, based on the growth of fluctuations seeded by
quantum fluctuations in the inflationary era \cite{MukFelBra92,KodSas84}, is one of the great success stories in cosmology (\cite{Lon13}:
Chapter 13). A key realisation was that introduction of a cosmological
constant would allow a cold dark matter scenario to match observations of an
almost flat universe \cite{EfsSutMad90}.
\index{FLRW models!perturbations|)}

\subsection{Precision cosmology: Cosmological success and puzzles}
\index{precision cosmology}

Use of the perturbed FLRW models in conjunction with extraordinary
observational advances has led to an era of precision cosmology, combining
physics and astrophysics with general relativity to predict both the
background model evolution plus the growth of structure in it. Key features
are: the dynamical evolution of the background model, with an inflationary
era followed by a hot big bang era, a matter dominated era, and a dark
energy dominated late era; physical modelling of the hot big bang era,
including nucleosynthesis\index{big bang nucleosynthesis} and matter-radiation decoupling leading to CMB
with a precise black body spectrum; models of galaxy and other source
numbers with respect to distance as well as luminosity versus distance
relations; modelling of structure formation, related to matter power spectra
and Baryon Acoustic Oscillations; and prediction of CMB angular power
spectra. Together they comprise the concordance model of cosmology
\cite{Dod03,Muk05,EllMaaMac12,PetUza13}\index{concordance model}.

The puzzles are that we do not know the nature of the inflaton\index{inflaton}, for which
there are over 100 models \cite{MarRinVen13}; we do not know the nature of
the dark matter that is indicated to exist by dynamical studies, and is far
more abundant than baryonic matter; and we do not know the nature of the
dark energy causing an acceleration of the expansion of the universe at late
times.%
\index{acceleration of the Universe} %
 It is possible that some of these issues may be indicating we need a
different theory of gravity than general relativity, for example MOND or a
scalar-tensor theory.

\subsection{More General Geometries: LTB and Bianchi Models}

There are two other classes of exact solutions that have been used quite
widely in cosmological studies.

\subsubsection{LTB spherically symmetric models}

Firstly, the growth of inhomogeneities may be studied by using exact
spherically symmetric solutions, enabling study of non-linear dynamics. The
zero pressure such models are the Lema\^{\i}tre-Tolman-Bondi exact solutions 
\cite{Bon47,Kra06} where the time evolution of spherically symmetric
shells of matter is governed by a radially dependent Friedmann equation. The
solutions generically have a matter density that is radially dependent, as
well as a spatially varying bang time. These models can be used to study (i)
the way a spherical mass with low enough kinetic energy breaks free from the
overall cosmic expansion and recollapses, hence putting limits on the rate
of growth of inhomogeneities \cite{Bon56}, and (ii) the way that any
observed $(m,z)$ and $(N,z)$ relations can be obtained in a suitable LTB
model where one runs the EFE backwards to determine the free functions in
the metric from observations, for any value whatever of the cosmological
constant $\Lambda $ \cite{MusHelEll99,EllMaaMac12}$.$ This opens up
the possibility of doing away with the need for dark energy if we live in an
inhomogeneous universe model, where the data usually taken to indicate a
change of expansion rate in time due to dark energy are in fact due to a
variation of expansion rate in space. However although the supernova
observations can be explained in this way, detailed observational studies
based in the kinematic Sunyaev-Zeldovich effect show this is unlikely \cite%
{EllMaaMac12}.

\subsubsection{The Bianchi models and phase planes}
\index{Bianchi models}

Secondly, following the work of G\"{o}del, Taub, and Heckmann and Sch\"{u}%
cking, there is a large literature examining the properties of spatially
homogeneous anisotropically expanding models. These models are generically
invariant under 3-dimensional continuous Lie groups of symmetries, whose Lie
algebra was first investigated by Luigi Bianchi using projective geometry;
much simpler methods are now available \cite{EllMac69}. Special cases allow
higher symmetries (Locally Rotationally Symmetric models \cite{Ell67}).\
These allow a rich variety of non-linear behaviour, including anisotropic
expansion at early and late times even if the present behaviour is nearly
isotropic; different expansion rates at the time of nucleosynthesis than in
FLRW models, leading to different primordial element abundances; complex
anisotropy patterns in the CMB sky; and much more complex singularity
behaviour than in FLRW models, including cigar singularities, pancake
singularities (where particle horizons may be broken in specific
directions), chaotic (`mixmaster') type behaviour \cite{Mis69}
characterised by `billiard ball' dynamics, and non-scalar singularities if
the models are tilted. Dynamical systems methods can be used to show the
dynamical behaviour of solutions and the relations of families of such
models to each other \cite{WaiEll97}.  If a cosmological model is generic,
it should include Bianchi anisotropic modes as well as inhomogeneous modes,
and may well show mixmaster behaviour \cite{LifKha64}.

\subsection{Cosmological success and puzzles}

Standard cosmology is a major application of GR showing how matter curves
spacetime and spacetime determines the motion of matter and radiation. It is
both a major success, showing how the dynamical nature of spacetime
underlies the evolution of the universe itself, with this theory tested by a
plethora of observations \cite{EllMaaMac12}, and a puzzle, with three major
elements of the standard model unknown (Dark matter, Dark energy, the
Inflaton). While much of cosmological theory (the epoch since decoupling)
follows from Newtonian gravitational theory (NGT), this is not true of the
dynamics of the early universe, where pressure plays a key role in
gravitational attraction: thus for example NGT cannot give the correct
results for nucleosynthesis. The theory provides a coherent view of
structure formation (with a few puzzles), and hence of how galaxies come
into existence. The theory raises the issue that the universe not only
evolves but (at least classically) had a beginning, whose dynamics lies
outside the scope of standard physics because it lies outside of space and
time. This is all discussed in depth in Chapter 4.


\section{Gravitational Lensing and Dark Matter}
\index{gravitational lensing|(}

\label{lens} 
The Schwarzschild solution predicts bending of light by massive objects:
that is, gravitational lensing will occur. Indeed any mass concentration
will differentially attract light rays; this is nowadays an important \
effect in astronomy and cosmology.

The bending of light by a gravitational field was predicted by Einstein in
1911 from the equivalence of a uniform gravitational field with an
accelerated reference frame. In 1912 he derived an equation for
gravitational lensing \cite{RenSauSta97}, well before he deduced the
gravitational field equations. In 1915, he applied the gravitational field
equations and found the deflection angle is twice the result obtained from
the equivalence principle, the factor two arising because of the curvature
of space. He predicted bending of light by the Sun in 1916, as famously
verified by Eddington's Expedition in 1919, but he only published his
pioneering paper on the bending of light by stars and galaxies in 1936 \cite%
{Ein36}. He stated there that the bending would not be observable in these
cases because of the small angular scales involved, but recognised that
lensing would cause fluctuations in the brightness of distant objects that
might be detectable.\newline

Gravitational lensing is now a key part of astronomy and cosmology \cite%
{SchEhlFal92,Wam01}.

\subsection{Calculating lensing}

One can calculate deflection of light, and associated image displacement,
distortion, and brightening, either through a weak field approximation,
which will be valid for any mass distribution, or by determining geodesics
in an exact solution, for example a spherically symmetric solution. They of
course have to agree in the weak field approximation. Both show that for
weak lensing, the angle of deflection of light caused by a spherical mass $M$
is: 
\[
\mathbf{\hat{\alpha}}(\mathbf{\xi })=\frac{4GM}{c^{2}}\frac{\mathbf{\xi }}{%
|\mathbf{\xi }|^{\mathbf{2}}} 
\]%
where $\mathbf{\xi }$ is the impact parameter, that is, the closest
distance the light beam would reach from the centre of the lensing mass if
there were no bending. Because the Schwarzschild radius of the Sun is $2.95$
km and the solar radius is $6.96\times 10^{5}$ km, a light ray grazing the
limb of the Sun is deflected by an angle ($5.9/7.0)\times 10^{-5}$ radians 
$=1.$"$7$.

For more complex mass distributions, we represent the mass projected onto a
mass sheet called the lens plane. Following Blandford and Narayan \cite%
{BlaNar92}, a light ray from a source $S$ at redshift $z$, is incident on a
deflector or lens L at redshift $z_{d}$ with impact parameter $\xi $
relative to some fiducial lens center. Assuming the lens is thin compared to
the total path length, its influence can be described by a deflection angle $%
\hat{\alpha}(\xi )$ (a two-vector) for the ray when it crosses the lens
plane. The mass projected onto this plane is characterized by its surface
mass density
\[
\Sigma (\mathbf{\xi })=\int \rho (\mathbf{\xi },z)dz 
\]%
where $\mathbf{\xi }$ is a two-dimensional vector in the lens plane. The
deflection angle at position $\mathbf{\xi }$ is the sum of the deflections
due to all the mass elements in the plane:
\[
\mathbf{\hat{\alpha}}(\mathbf{\xi })=\frac{4\pi G}{c^{2}}\int \frac{(\mathbf{%
\xi }-\mathbf{\xi }\prime )\Sigma (\mathbf{\xi }\prime )}{|\mathbf{\xi }-%
\mathbf{\xi }\prime |^{2}}d^{2}\mathbf{\xi }\prime .\ 
\]%
In the cosmological context, the intrinsic lensing angle $\mathbf{\hat{\alpha%
}(\theta })$ must be corrected to give the observed lensing angle $\mathbf{%
\alpha }$ by using the angular diameter distances $D_{d}$, $D_{s}$, and
$D_{ds}$
between the source, deflector, and observer, which are affected by
cosmological parameters. When the deflected ray reaches the observer, she
sees the image of the source apparently at position $\theta $ on the sky.
The true direction of the source, i.e. its position on the sky in the
absence of the lens, is given by $\beta $. Now $\theta D_{s}=\beta D_{s}-%
\mathbf{\alpha }D_{ds}$ where $\beta $ is the angle from the source centre
to the lensing element. Therefore, the positions of the source and the image
are related through the lensing equation

\[
\mathbf{\beta }=\mathbf{\theta }-\mathbf{\alpha }(\mathbf{\theta })
\]%
where $\mathbf{\alpha }(\mathbf{\theta })=\frac{D_{ds}}{D_{s}}\mathbf{\hat{%
\alpha}(}D_{\mathbf{d}}\mathbf{\theta }).$ This light deflection leads to
image distortion and amplification, characterised by the image magnification
and shear. The shear is caused by the Weyl tensor the light encounters; the
magnification is caused directly by the matter it encounters, and indirectly
by the cumulative effect of the shear \cite{KriSac66,NewPen62,HawEll73} (this follows from (1.5) with (1.8), and leads to (1.7)). The
more mass (and the closer to the center of mass), the more the light is
bent, and the more the image of a distant object is displaced, distorted,
and perhaps magnified. These are the basic equations from which the
cosmological applications follow.

\subsection{Strong lensing}

Lensing by galaxies can bend the light of a point background source by a
large enough angle that it can be observed as several separate images. If
the source, lensing object, and observer lie in a straight line, the source
can appear as a ring around the lensing object (this is the case of Einstein
rings). Misalignment will result in an arc segment instead. Lensing masses
such as galaxy groups and clusters can result in the source being seen as
many arcs around the lens. The observer may then see multiple distorted
images of the same source, with time delays detectable between them if the
source varies. The time delay between images is proportional to the
difference in the absolute lengths of the light paths, which in turn is
proportional to $H_{0}$. The number and shape of the arcs depends on the
relative positions of the source, lens, and the observer, as well as the
gravitational well of the lensing object. The number and position of images
is due to caustics in the light sheet that can be characterised as cusps,
folds, etc. on using catastrophe theory \cite{Per04}.  Details are given
in 
Chapter 3.

This effect enables detecting very distant galaxies that are otherwise
unobservable. For example combining observations from the Hubble Space
Telescope, Spitzer Space Telescope, and gravitational lensing by cluster
Abell 2218, allowed discovery of the galaxy MACS0647-JD, that is roughly 13
billion light-years away. The galaxy appears 20 times larger and over three
times brighter than typically lensed galaxies. 

\subsection{Weak lensing}

Weak lensing results in image distortion, but the shear may be too small to
be seen directly. Also apparent shear may be due to the galaxy's distinct
shape, an angle of view that makes it appear elongated, or may be due to the
telescope, the detector, or the atmosphere, so one cannot deduce weak
lensing from images of a single object. However the faint distortions due to
lensing of images of a set of galaxies can be worked out statistically, and
the average shear due to lensing by some massive objects in front can be
computed. Thus in the weak lensing case, measuring statistical distortion is
the key to measuring the mass of the lensing object. Weak lensing surveys
use this method to determine the intervening mass distribution. This is an
important method in detecting dark matter \cite{SchEhlFal92}. Indeed the
most important result from weak lensing concerns the proof of existence of
collionless dark matter in the `bullet cluster' (see Chapter
3).
An exciting
recent application is detection of weak lensing in observations of CMB\
anisotropy patterns by the Planck satellite.

\subsection{Microlensing}

For point objects, it may happen that no distortion in shape can be seen but
the amount of light received from a distant object changes in time, with a
characteristic shape of the resulting light curve. The lensing objects may
be stars in the Milky Way, with the source being stars in a remote galaxy,
or a distant quasar. The method has been used to discover extrasolar
planets, which is one of its most important applications.

\subsection{Conclusion}

Gravitational lensing was predicted very early on by Einstein and provided
the first new observational evidence that vindicated general relativity. It
is a confirmation of the curving of spacetime by matter and allows us to
detect matter which has no other observational signature. It therefore plays
a key role in the mapping of dark energy in the universe. It also acts a
lens allowing us to see far further than otherwise possible. It is thus a
key tool in cosmology, discussed in the chapter 3. 
\index{gravitational lensing|)}

\section{Gravitational Waves}
\index{gravitational waves|(}

\label{gravwaves} 
Einstein also predicted the existence of gravitational waves very early on 
\cite{Ein18}. However because of the coordinate freedom of general
relativity, there was confusion as to whether these were real waves, or just
coordinate waves. This was sorted out by the understanding, particularly due
to Pirani, that the gravitational waves could be seen as Weyl tensor waves 
\cite{Pir57}, which could carry away energy and momentum. They could
therefore be indirectly observed by their effect on binary pulsar orbits,
which can be measured to extremely high precision. Direct detection is much
more difficult, but spectacular development in detector design promises that
they may be detected in the next decades, with gravitational wave
observatories having the potential to become an essential tool in precision
cosmology. Crucial to this project are major developments in numerical
relativity allowing us to predict the nature of waves expected to be emitted
from binary blackhole mergers and other strong field sources.

\subsection{Weak field formulation}
\index{gravitational waves!weak}

Gravitational waves exist because GR is a relativistic theory of
gravitation; unlike Newtonian theory, influences cannot be exerted
instantaneously. Using a weak field approximation $g_{%
{\mu}%
\nu }=\eta _{%
{\mu}%
\nu }+h_{%
{\mu}%
\nu },$ $h_{%
{\mu}%
\nu }<<1$ , and harmonic coordinates, Einstein predicted existence of waves
in the metric tensor. But this may just be a wave in the coordinates! One
can avoid this problem by using geometric variables, for example the 1+3
covariant representation of the Weyl tensor, whereby one obtains Maxwell
like equations for the electric and magnetic parts of the Weyl tensor \cite%
{Haw66,Ell71,MaaBas98}.

Nevertheless usual calculations use the weak field method of Einstein \cite%
{Ein18}. The perturbation $h_{ab}$ is tracefree and transverse, so
gravitational waves are transverse and, given their direction of propagation 
$k$, have two degrees of freedom (two polarisations) as represented by trace
free 2-tensors orthogonal to $k.$ That is, they are spin-2 fields. Defining $%
\bar{h}_{ab}\equiv h_{ab}-\frac{1}{2}\eta _{ab}h\ $and choosing coordinates
so that $\bar{h}_{\;\;,b}^{ab}=0,$ then on defining $z^\prime =z-ct,$ the two
polarisation modes are $h_{+}(z\prime )$ and $h_{-}(z^\prime ),$ with weak
field metric form

\begin{eqnarray*}
ds^{2} &=&-dt^{2}+dx^{2}+dy^{2}+dz^{2}\;+\ h_{\alpha \beta }(z\prime
)dx^{\alpha }dx^{\beta }\;(h\ll 1), \\
h_{ab} &=&\left( 
\begin{tabular}{lll}
0 & 0 & 0 \\ 
0 & $h_{+}$ & $h_{-}$ \\ 
0 & $h_{-}$ & -$h_{+}$%
\end{tabular}%
\right) .
\end{eqnarray*}%
One can get plane wave solutions of this form, characterised by their
amplitude $h,$  frequency $f$, wavelength $\lambda $, and speed $c$,
related by the usual equation $c=\lambda f$. They can be linearly polarized,
with quadrupole polarization components $h_{+}$ and $h_{\times }$ in
canonical coordinates, rotated by 45 degrees relative to each other. One can
also have circularly polarized waves.

In the slow motion approximation for a weak metric perturbation $\ $for a
source at distance $r,$ the wave radiated by a local source is given by

\[
h_{%
{\mu}%
\nu }=\frac{2G}{c^{4}r}\frac{d^{2}I_{%
{\mu}%
\nu }}{dt^{2}} 
\]%
where $I_{%
{\mu}%
\nu }$ is the reduced quadrupole moment defined as

\[
I_{%
{\mu}%
\nu }=\int \rho (\mathbf{r})\left( x_{%
{\mu}%
}x_{\nu }-\frac{1}{3}\delta _{%
{\mu}%
\nu }r^{2}\right) dV. 
\]%
Hence gravitational waves are radiated by objects whose motion involves
acceleration, provided that the motion is not perfectly spherically
symmetric (the case of an expanding or contracting sphere). There are no
spherical gravitational waves! An example is two gravitating objects that
form a binary system. Seen from the plane of their orbits, the quadrupole
term $I_{%
{\mu}%
\nu }$ will have non-zero second derivative.

\subsection{Exact Gravitational waves}

One can find exact gravitational wave solutions for cases with high
symmetries.

\subsubsection{Cylindrical gravitational waves}
\index{gravitational waves!cylindrical}

Einstein and Rosen \cite{EinRos37} derived the exact solution for cylindrical
gravitational waves. They state, \textquotedblleft The rigorous solution for
cylindrical gravitational waves is given. After encountering relationships
which cast doubt on the existence of rigorous solutions for undulatory
gravitational fields, we investigate rigorously the case of cylindrical \
gravitational waves. It turns out that rigorous solutions exist and that the
 problem reduces to the usual cylindrical waves in euclidean
space.\textquotedblright\ 

\subsubsection{Plane gravitational waves}
\index{gravitational waves!plane}

Plane gravitational waves are Petrov type N vacuum solutions that allow
arbitrary information to be carried at the speed of light. They are exact
solutions of the empty space field equations that can have high symmetries
as studied by Ehlers and Kundt, and by Bondi, Pirani, and Robinson \cite%
{BonPirRob59}. They have intriguing global properties due to the way they focus
null geodesics \cite{Pen65a}. One can find exact solutions for two colliding
plane gravitational waves \cite{Sze72}.

\subsection{Asymptotic flatness and gravitational radiation}
\index{asymptotic flatness}\index{gravitational radiation}

Gravitational waves can carry energy, momentum, and information, and it is
useful to investigate this in general cases without symmetry. This can be
done by asymptotic expansions in flat spacetimes, using multipole expansions
clarified nicely by Thorne 
\cite{Tho80}. Following Trautman, outgoing
radiation conditions have been formulated by Bondi, Sachs, Newman and
Penrose, and others, leading to peeling-off theorems showing how
successively different Petrov types occur at larger distances, and a `news
function' related to the derivative of shear is related to mass loss
(\cite{BonvanMet62,Sac62,NewPen62}). Penrose showed how to express
all this using his conformal representation of infinity \cite{PenRin84}.
Positive mass theorems aim to show that one cannot radiate so much mass away
that the mass becomes negative \cite{SchYau79,Wit81}.

\subsection{Emission}
\index{gravitational waves!emission}\index{gravitational
waves!emission|see{Chapter 6}}

The linear perturbation formulae given above suffice for gravitational wave
emission in weak field situations like the binary pulsar, and there is a
large literature on analytic perturbations of black hole solutions \cite%
{Teu73} and the black hole perturbation approach to gravitational radiation 
\cite{SasTag03,KokSch99} as well as studies of gravitational waves
from gravitational collapse \cite{FryNew11} and from post-Newtonian sources
and inspiralling compact binaries \cite{Bla06}. The Post-Newtonian (PN) approximation turned out to work even
better than expected for inspiralling compact sources and it provides a
 very accurate and physical picture (including spins, eccentric orbits,
tidal effects, etc). On the other hand key progress
has come from major developments in numerical methods for examining non-linear relativity (NR) 
processes such as coalescence of two black holes, and associated numerical
codes \cite{OweBriChe11,NicOweZha11,ZhaZimNic12,NicZimChe12}. Progress in this area
has particularly been due to breakthroughs by Frans Pretorius. \\

These methods are complementary. The PN cannot describe the merger of two
BHs, on the other hand NR will not be able to compute the inspiral of two BH or NS as does the PN, simply because of the prohibitive computing time
 needed to control 20,000 cycles before the coalescence, and the extreme
 accuracy of the PN at large separations.
 In LIGO/VIRGO dectectors the PN plays a crucial role, both for detecting
 the signals on-line and for subsequent analysis and parameter estimation
off-line. For sources like coalescing neutron stars, the templates are in
 fact completely based on the PN methods. Even for BH binaries with masses up to
 say 20 solar masses, the PN waveform constitutes a large part of the
 templates. Idem for future supermassive BHs sources for LISA.
The theoretical prediction from GR is a
 combination of PN for the inspiral and NR for the merger, both being
 accurately matched together.\\

Chapter 6 discusses the PN methods, and Chapter 7
discusses probing strong field gravity through numerical simulations.

\subsection{Gravitational wave detection \label{sec:gravwavecosm}}
\index{gravitational waves!detection}\index{gravitational
waves!detection|see{Chapter 5}}

As shown by Pirani, the effect of gravitational waves on local systems is
via the (generalised) geodesic deviation equation \cite{Pir56,Pir57},
which shows how the wave will tend to move particles transverse to
the direction of motion in a quadrupole mode. Joseph Weber pioneered the
effort to build gravitational detectors based on this effect, but it is
enormously difficult technically because the strain is so small: typically
of amplitude $h\simeq 10^{-20}.$ Indirect detection via the effects of
energy loss is easier.

\subsubsection{Indirect detection: Binary pulsars}
\index{binary pulsar}

As noted above, emission requires anisotropic motion of a mass. This occurs
in astronomical binary systems. If two masses $m_{1}$ and $m_{2}$ are in
orbit around each other, separated by a distance $r$, the power radiated by
this system is:

\begin{equation}
P=\frac{dE}{dt}=-\frac{32}{5}\frac{G^{4}}{c^{5}}\frac{%
(m_{1}m_{2})^{2}(m_{1}+m_{2})}{r^{5}}
\end{equation}%
This energy loss will be reflected in a change in the orbital period, which
is measurable in the case of binary pulsars because their precise pulse
timing allows very accurate orbital tracking. Observations of binary pulsar
decay rates by Hulse and Taylor confirmed this effect, and so confirmed the
existence of gravitational radiation carrying energy away from the system 
\cite{TayFowMcC79,TayWei89}.

\subsubsection{Indirect Detection: Inflation}
\index{inflation}

Gravitational waves are expected to be emitted in the inflationary era in
the very early universe and so CMB observations can provide evidence for
their existence \cite{TurWhiLid93}. The relation between scalar contributions $%
S$ to the quadrupole and tensor contributions $T$ determines the amplitude $%
(T/S)$. This is related to the tilt $n_{T}$ by the key relation 
\begin{equation}
n_{T}=-\frac{1}{7}\frac{T}{S}
\end{equation}%
which not only provides a consistency check of inflation, but it
allows direct detection of gravity waves, as it relates the overall
amplitude to the tilt \cite{Tur96}. The ratio $T/S$ must be greater
than $0.2$ for a statistically significant detection of tensor
perturbations. Gravitational waves will imprint as B-modes in the CMB\
anisotropy patterns, and so should be detectable by CMB polarisation
measurements (\cite{BonEfs84,Dod03}). The value of the tensor to
scalar ratio determined by observations such as those by Planck and BICEP2
strongly constrain models of inflation \cite{MarRinTro14}. However foreground effects have to be very  carefully controlled, and have undermined the original BICEP2 results.

\subsubsection{Direct Detection by bars or interferometers}

Ground-based direct detection is possible in principle via bar detectors or
interferometers. Because of the very small size of the strain, this requires
immense skill in technical development, discussed in Chapter 5; in
particular it requires quantum non-demolition methods pioneered by
Braginski. Gravitational-wave experiments with interferometers and with
resonant masses can search for stochastic backgrounds of gravitational waves
of cosmological origin. The sensitivity of these detectors as a function of
frequency has been carefully explored in relation to the expected
astronomical sources, and this method has the prospect of opening up a new
window on the universe because it will reach back to much earlier times than
any other method \cite{KraDodMey10,SatSch09}. Direct detection of
the inflationary gravitational wave background constrains inflationary
parameters and complements CMB polarization measurements \cite{KurGorSil10}.

\subsubsection{Pulsar timing arrays}

Gravitational waves will affect the time a pulse takes to travel from a
pulsar to the Earth. A pulsar timing array uses millisecond pulsars to
search for effects of gravitational waves on measurements of pulse arrival
times, giving delays of less than $10^{-6}$ seconds. Three pulsar timing
array projects are searching for patterns of correlation and
anti-correlation between signals from an array of pulsars that will signal
the effects of gravitational waves.

\subsection{Conclusion}

The theory of gravitational waves extends the idea of transverse wave
propagation from the spin-1 field of electromagnetism (Maxwell's theory) to
the spin 2 field of gravitation. They are an essentially relativistic
phenomenon: they cannot occur in Newtonian gravitational theory.
Gravitational waves can carry energy and arbitrary information, and indeed
convey information to us from the earliest history of the universe that we
can access. They provide the ultimate limit of our possible access to
knowledge about the early universe. Their direct detection is a formidable
technological problem: the detectors being constructed are a triumph of
theory realised in practice. Chapter 6 discusses sources of gravitational
waves: theory and observations while Chapter 5 discusses current and future
ground and space based laser interferometric gravitational wave
observatories, pulsar timing arrays and the Einstein Telescope. It
summarizes how these efforts will provide a brand new window on the universe.
\index{gravitational waves|)}

\section{Generalisations}

Classical generalisations of general relativity include scalar tensor
theories, higher derivative theories, theories with torsion, bimetric
theories, unimodular theories, and higher dimensional theories. However if
one demands only second order equations in four dimensions and with one
spacetime metric, general relativity is the unique gravitational theory
based in Riemannian geometry, as shown by Lovelock \cite{Lov71}. The theory
was derived not because of experiment, but as the result of pure thought;
but it has survived all experimental tests \cite{Wil06}: see Chapter 2.
There is no
observational or experimental reason to modify or abandon the field
equations.

However there is a significant problem in terms of the relation of
general relativity to quantum field theory calculations that predict
existence of a vacuum energy density vastly greater than the observed
value of the cosmological constant \cite{Wei89}. Possible solutions
include either the existence of a multiverse combined with
observational selection effects, or some form of unimodular gravity
leading to the trace free form of the Einstein equations. In the
latter case the exactly constant vacuum energy density does not
gravitate, and this major problem is fully solved, with no change to
all the results mentioned above in this chapter.

General relativity theory, and all its applications (most of which he
pioneered), are yet another testament to Albert Einstein's extraordinary
creativity and physical insight.

\end{document}